%% file: ms.tex
\documentclass[11pt,a4paper]{article}
\pdfoutput=1 

\usepackage{jcappub}
\usepackage{subcaption}
\usepackage{float}
\usepackage{xfrac} 
\usepackage{indentfirst}
\usepackage{graphicx,amssymb}
\usepackage{url}

\def\be{\begin{equation}}
\def\ee{\end{equation}}
\def\ba{\begin{align}}
\def\ea{\end{align}}

\def\lsim{\raise0.3ex\hbox{$\;<$\kern-0.75em\raise-1.1ex\hbox{$\sim\;$}}}
\def\gsim{\raise0.3ex\hbox{$\;>$\kern-0.75em\raise-1.1ex\hbox{$\sim\;$}}}
\def\e{{\mathrm e}}
\def\epsilon{\varepsilon}
\def\theta{\vartheta}

\def\lsim{\raise0.3ex\hbox{$\;<$\kern-0.75em\raise-1.1ex\hbox{$\sim\;$}}}
\def\gsim{\raise0.3ex\hbox{$\;>$\kern-0.75em\raise-1.1ex\hbox{$\sim\;$}}}

\def\Rpos{f_{e^+/e^-}}

\renewcommand{\vec}[1]{\boldsymbol{#1}}
\renewcommand{\d}{\mathrm d}

\font\menge=bbold9 scaled \magstep1
\def\nota#1{\hbox{$#1\textfont1=\menge $}}
\def\MR{\nota R}

\def\hide#1{}

\title{Cosmic ray positrons from compact binary millisecond pulsars}

\author[a,b,c]{M.~Linares}
\author[c]{and M.~Kachelrie{\ss}}

\affiliation[a]{Departament de F{\'i}sica, EEBE, Universitat Polit{\`e}cnica de Catalunya,\\ Av. Eduard Maristany 16, E-08019 Barcelona, Spain.}
\affiliation[b]{Institute of Space Studies of Catalonia (IEEC), E-08034 Barcelona, Spain.}
\affiliation[c]{Institutt for fysikk, NTNU, Trondheim, Norway.}

\abstract{A new population of neutron stars has emerged during the
last decade: compact binary millisecond pulsars (CBMSPs).
Because these pulsars and their companion stars are in tight orbits
with typical separations of $10^{11}$\,cm, their winds interact strongly
forming an intrabinary shock.
Electron-positron pairs reaccelerated at the shock can reach
energies of about 10~TeV, which makes this new population a
potential source of GeV-TeV cosmic ray positrons.
We present an analytical model for the fluxes and spectra of positrons
from intrabinary shocks of CBMSPs.
We find that the minimum energy $E_{\min}$ of the pairs that enter the
shock is critical to quantify the energy spectrum with which positrons
are injected into the interstellar medium.
We measure for the first time the Galactic scale height of CBMSPs, 
$z_\e=0.4\pm 0.1$\,kpc, after correcting for an observational bias against 
finding them close to the Galactic plane.
From this, we estimate a local density of 5--9\,kpc$^{-3}$ and an
extrapolated total of 2--7 thousand CBMSPs in the Galaxy.
We then propagate the pairs in the isotropic diffusion approximation
and find that the positron flux from the total population is about two times
higher than that from the 52~currently known systems.
For $E_{\min}$  between 1 and 50\,GeV, our model predicts only a minor 
contribution from CBMSPs to the diffuse positron flux at 100\,GeV observed 
at Earth.
We also quantify the effects of anisotropic transport due to the ordered Galactic 
magnetic field, which can change the diffuse flux from nearby sources drastically.
Finally, we find that a single ``hidden" CBMSP close to the Galactic
plane can yield a positron flux comparable to the AMS-02 measurements
at 600\,GeV if its line-of-sight to Earth is along the ordered
Galactic field lines, while its combined electron and positron flux at
higher energies would be close to the measurements of CALET, DAMPE and
Fermi-LAT.  }

\date{\today}


\begin{document}

\maketitle

\section{Introduction}
\label{sec:intro}

\subsection{Cosmic ray positrons and pulsar winds}
\label{sec:intro-cray}

Measurements of the antimatter fraction of cosmic rays (CRs) are not
only valuable probes for cosmology and particle physics, but provide
also important insights into astrophysical sources of CRs and their subsequent
propagation in the Galaxy~\cite{Kachelriess19}.
A guaranteed production channel of antimatter are the interactions of
CR protons and nuclei with gas in the interstellar medium.
If this channel were the main source of antimatter, the energy dependence
of the Galactic diffusion coefficient ($D\propto E^\delta$ with
$\delta=0.3-0.5$) would lead to a decrease of the
antimatter fraction with energy, as discussed e.g.\ in
Ref.~\cite{Serpico12}
The rise in the positron-to-electron fraction above 20\,GeV, first observed
clearly by the PAMELA~\cite{Adriani:2008zr,Adriani:2010rc} and then confirmed
by the AMS-02 collaboration~\cite{Aguilar:2014mma,Aguilar:2019owu}, requires
therefore additional sources of positrons.
Moreover, the high-energy part of the $e^\pm$ spectrum should be
dominated by local sources, as pointed out already 30~years
ago~\cite{1987ICRC....2...92H,1989ApJ...342..807B}, since high-energy
electrons lose energy fast.

Pulsars are natural candidates for such positron sources,
because electromagnetic pair cascades in their magnetospheres lead to a
large positron fraction, $\Rpos\simeq 1$, without affecting the
antiproton flux.
A neutron star with spin period $P$, radius $R$, and surface magnetic
field strength $B_\mathrm{s}$ must maintain a charge density known as
the Goldreich-Julian density~\cite{GJ69} to screen the electric
field induced by its rotating magnetic field.
In order to keep this charge density at a steady state, charged
particles are extracted from the neutron star surface and ejected at a
rate that can be characterized by the Goldreich-Julian
rate\footnote{We use Gaussian cgs units throughout this work unless stated otherwise.}  \cite{Kirk09},
\begin{equation}
\label{eq:Ngj}
\dot{N}_\mathrm{GJ} = \frac{4 \pi^2 B_\mathrm{s} R^3}{2 c e P^2} ,
\end{equation}
of order 10$^{32}/$s for the pulsars studied here.
According to most models~\cite{Kirk09,Harding11}, these
``primary'' charges lead to copious pair production inside the pulsar
magnetosphere, within the light cylinder defined as the surface where
corotating magnetic field lines reach the speed of light $c$, at a
radius $R_\mathrm{lc} = cP/(2\pi)$.
Depending on the pair cascade multiplicity, the rate of ``secondary'' pairs
emerging from the light cylinder
can be 1--4 orders of magnitude higher than $\dot{N}_\mathrm{GJ}$.

These secondary $e^\pm$ pairs escape through the open field lines
and feed the relativistic pulsar wind.
The contribution of protons and nuclei to the wind and shock in an
additional hadronic component has been debated for decades~\cite{Arons94}.
Pulsar winds are thought to be Poynting flux dominated near the light
cylinder, in their innermost parts.
In other words, the electromagnetic energy density is thought to dominate
over the  kinetic energy density of particles just outside the light cylinder,
in the inner wind.
How and where particles are accelerated as one moves outwards in the
wind is an open question (the so-called ``sigma problem"; see, e.g.,
Ref.~\cite{Kirk09} for a recent review).
Reference~\cite{Bogovalov00} constrained the transition radius from a
magnetically to a kinetically dominated wind, placing it at more than
$(5-30)\,R_\mathrm{lc}$.
But at the distances of interest for our work, with intrabinary shocks
at $\sim (10^3-10^4)\,R_\mathrm{lc}$, the ratio of magnetic to particle
energy density is so far largely unconstrained.

The expected electron and positron fluxes from some pulsars, as well
as the resulting anisotropy, have been studied in detail using the
isotropic diffusion approximation~\cite{Hooper:2008kg,Grasso:2009ma}.
Two promising candidates are the relatively young, isolated, gamma-ray
pulsars Geminga and PSR~B0656+14, which are only (250--300)\,pc from
Earth. Recent HAWC observations of extended TeV gamma-ray emission
confirmed that they are local sources of accelerated
electrons~\cite{Abeysekara:2017old}.
However, the presence of electrons with energies in the 10--1000\,GeV
range in these sources can be tested directly looking for GeV photons
using {\it Fermi}'s Large Area Telescope (LAT).  No GeV photon halo
around Geminga and PSR~B0656+14 has been found in the search performed
in Ref.~\cite{Shao-Qiang:2018zla}, and the derived upper limits were
used to constrain their contribution to the observed positron flux as
$\lsim 15\%$.
A similar analysis~\cite{DiMauro:2019yvh} detected a weak GeV halo
around Geminga and set an upper limit of $20\%$ to its contribution to
the observed positron flux. 
These limits disfavour these two candidates, and young pulsars in
general, as explanation for the positron excess.

An alternative pulsar scenario are pulsar wind nebulae (PWN) with bow
shocks~\cite{Blasi:2010de,Bykov:2017xpo}. These shocks form when
PWNe move relative to the ambient interstellar medium with supersonic
speeds.
Particles accelerated at the termination surface of the pulsar
wind may undergo acceleration in the converging flow formed by the
outflow from the wind termination shock and the inflow from the bow
shock, leading to very hard energy spectra.
Assuming a steepening of the spectrum at $E\simeq 500$\,GeV, the
measured positron fraction can be reproduced.  Because pairs can only
be released into the interstellar medium after the pulsar has escaped
the supernova remnant ($\sim 10^4$~yr after being formed), the
predictions of this scenario depend critically on the time evolution
of the spin-down rate of the neutron star~\cite{Blasi:2010de}.
Specifically, for the most likely spin-down history the PWN bow shock
models require very high efficiencies ($\simeq 30-50\%$) in converting
spin-down luminosity into pairs~\cite{Blasi:2010de,Bykov:2017xpo}.

\subsection{Millisecond pulsars in compact binaries}
\label{sec:intro-spiders}

A new class of millisecond pulsars (MSPs) has emerged during the last
decade \cite{Roberts11,Hessels11,Ray12}, thanks to the {\it
  Fermi\/}-LAT.
These are nearby MSPs (mostly within 3\,kpc) in compact binaries
(orbital periods $P_\mathrm{b} \lesssim 1$\,d), with non-degenerate or
semi-degenerate companion stars.
The former, known as redbacks (RBs), have companions with masses of at
least $M_\mathrm{c,min} \sim 0.1\,M_\odot$~\cite{DAmico01}, while
the latter, known as black widows (BWs) have $M_\mathrm{c,min} \sim
0.01\,M_\odot$~\cite{Fruchter88}.
We refer to both types as compact binary MSPs (CBMSPs) or ``spiders",
after their nicknames inpired by cannibalistic spiders.
While in 2008 only four such systems were known, at the moment of writing
we know 52 CBMSPs in the Galactic field, so they represent about 20\%
of the total MSP population (see Section~\ref{sec:spiders-current} for
details).

Millisecond pulsars live long (their characteristic age is $P/(2\dot{P}) \sim
0.1-10$\,Gyr) and have moderately strong winds powered by the loss of
rotational energy with spin-down luminosities $L_\mathrm{sd} =
10^{33}-10^{35}$\,erg/s.
Since the magnetospheres of MSPs are smaller than those of normal
(young, slow) pulsars and that is where the magnetic field decays most
rapidly, MSPs have stronger fields than normal pulsars at the light
cylinder ($10^3-10^6$~G) and in the wind.
The orbital evolution of CBMSPs also happens on long timescales
($\gtrsim 0.5$\,Gyr), driven by the evaporation of the companion or
magnetic braking~\cite{Chen13,Ginzburg20}.
Since these are much longer than the CR diffusion timescales, CBMSPs
can be considered steady sources of CRs, as opposed to
quasi-instantaneous sources of CRs like young pulsars or supernova
remnants.

Due to their small orbital separations, $a \sim 10^{10}-10^{11}$\,cm,
the pulsar and its companion can interact strongly, providing
a new probe of the innermost pulsar wind.
The X-ray emission of CBMSPs indicates the presence of an
intrabinary shock between the pulsar and companion winds
\cite{Stappers03,Bogdanov10,Bogdanov11b}, which can be very efficient
at reaccelerating particles \cite{Harding90,Arons93}.
While pair cascades from the magnetospheres of MSPs cut off around a
few tens of GeV and thus cannot contribute to the high-energy rise of
$\Rpos$, electron-positrons are accelerated up to tens of TeV in the
strong intrabinary shocks of CBMSPs.
In particular, the authors of Ref.~\cite{Venter15} argued that the contribution
of 24~CBMSPs known in 2015 to the positron flux on Earth can reach levels of
a few tens of percent at tens of TeV, depending on model parameters.

In this work we examine if the full population of CBMSPs can
contribute significantly to the observed positron flux on Earth.
To do so, we study the currently known population and estimate the
total intrinsic Galactic population in Section~\ref{sec:spiders}.
In Section~\ref{sec:pairs}, we develop a simple analytical model of
the electron-positron fluxes and spectra from intrabinary shocks of
CBMSPs.  We calculate the diffuse positron flux on Earth including
effects of anisotropic diffusion (Section~\ref{sec:flux}) and present
the resulting diffuse fluxes in Section~\ref{sec:results}.
We briefly discuss and summarize our main results in Section~\ref{sec:summary}.

\input{source_table_clean.tex}

\section{A growing population of spiders}
\label{sec:spiders}

\subsection{Currently known population}
\label{sec:spiders-current}

Table~\ref{table:spiders} shows the currently known population of
CBMSPs and some of their properties, which we have collected from the
literature.
We do not include here globular cluster spiders, since most of them
are too distant to be relevant for our purposes (there are about 30
known CBMSPs in globular clusters, but all of them are more than 2\,kpc
away, and only four are between 2 and 4\,kpc~\cite{FreireCat}).
Our updated sample includes 52~systems: 22~redbacks (seven of them
candidates, where no pulsations have been detected yet, labelled RBc)
and 30~black widows (one of them candidate, labelled BWc).
The median values of $P$, $L_\mathrm{sd}$, $P_\mathrm{b}$ and
$M_\mathrm{c,min}$ are given in Table~\ref{table:spiders}.
In the few cases where these measurements are not available
(e.g.\ because no coherent pulsar timing solution has been published)
we use such median values as input for the pair spectral model
that we will develop in Section~\ref{sec:pairs}.

For those systems with reported spin period derivative $\dot{P}$, we
calculate the surface magnetic field strength at the equator as
\begin{equation}
\label{eq:Bs}
B_{\mathrm s} = 2 \sqrt{P \dot{P}\frac{3 I c^3}{8 \pi^2 R^6}}\,,
\end{equation}
where we assume $I=2MR^2/5$ for the neutron star's moment of inertia and we
use as
mass $M=1.85~M_{\odot}$~\cite{Linares19} and radius $R=11$\,km~\cite{Ozel16}.
We calculate the orbital separation,
\begin{equation}
a=\left(\frac{GM_\mathrm{tot}\,P_b^2}{4\pi^2}\right)^{(1/3)},
\end{equation}
assuming for the total mass in the binary $M_\mathrm{tot}=M+M_\mathrm{c,min}$.
We use refined distance measurements for each system when available (from,
e.g., radio parallax~\cite{Deller12} or optical observations~\cite{Kaplan13}),
and radio-timing dispersion measure estimates otherwise.

\begin{figure}[tbp]
\centering
\includegraphics[width=.48\textwidth,angle=0]{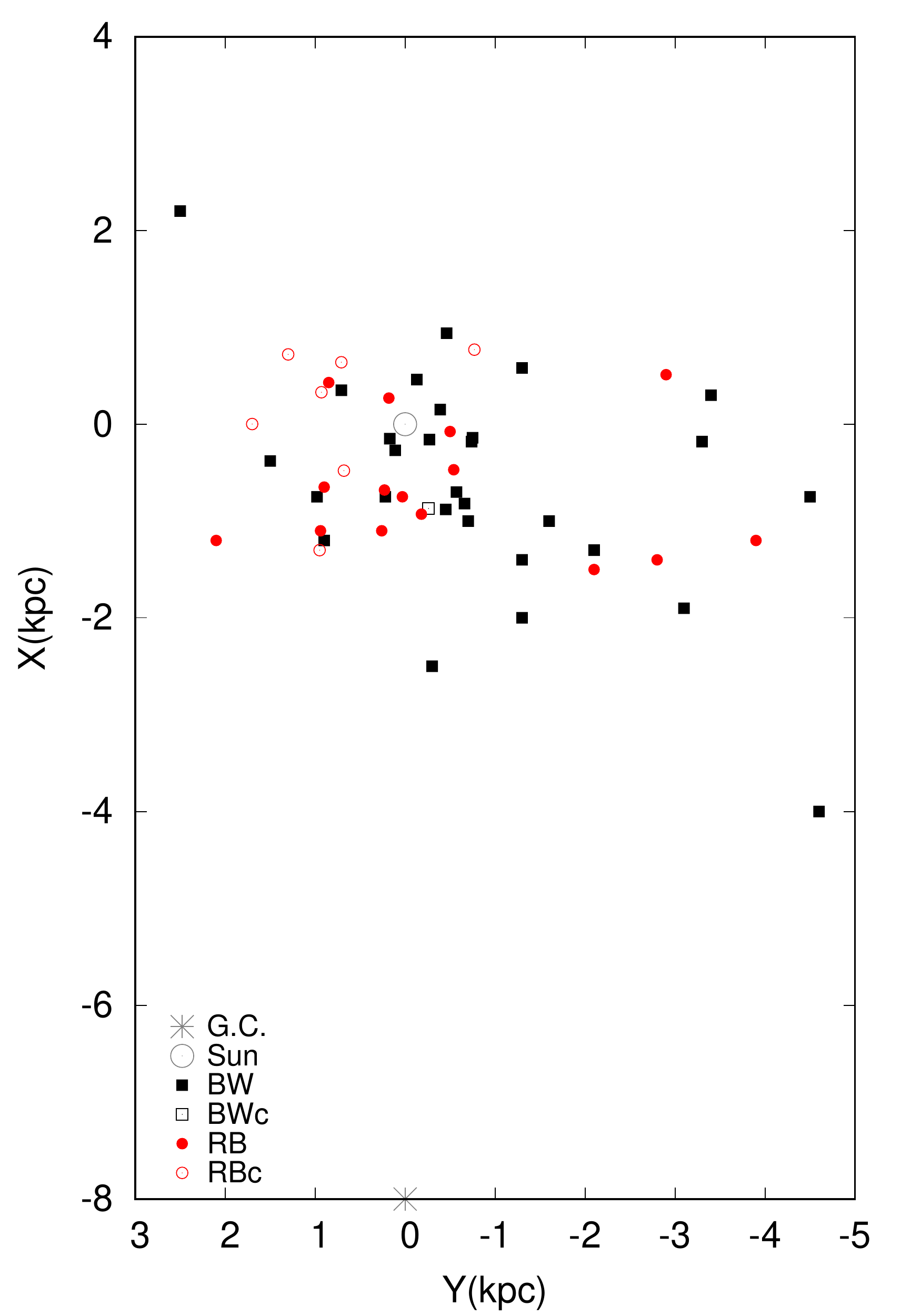}
\hfill
\includegraphics[width=.48\textwidth,angle=0]{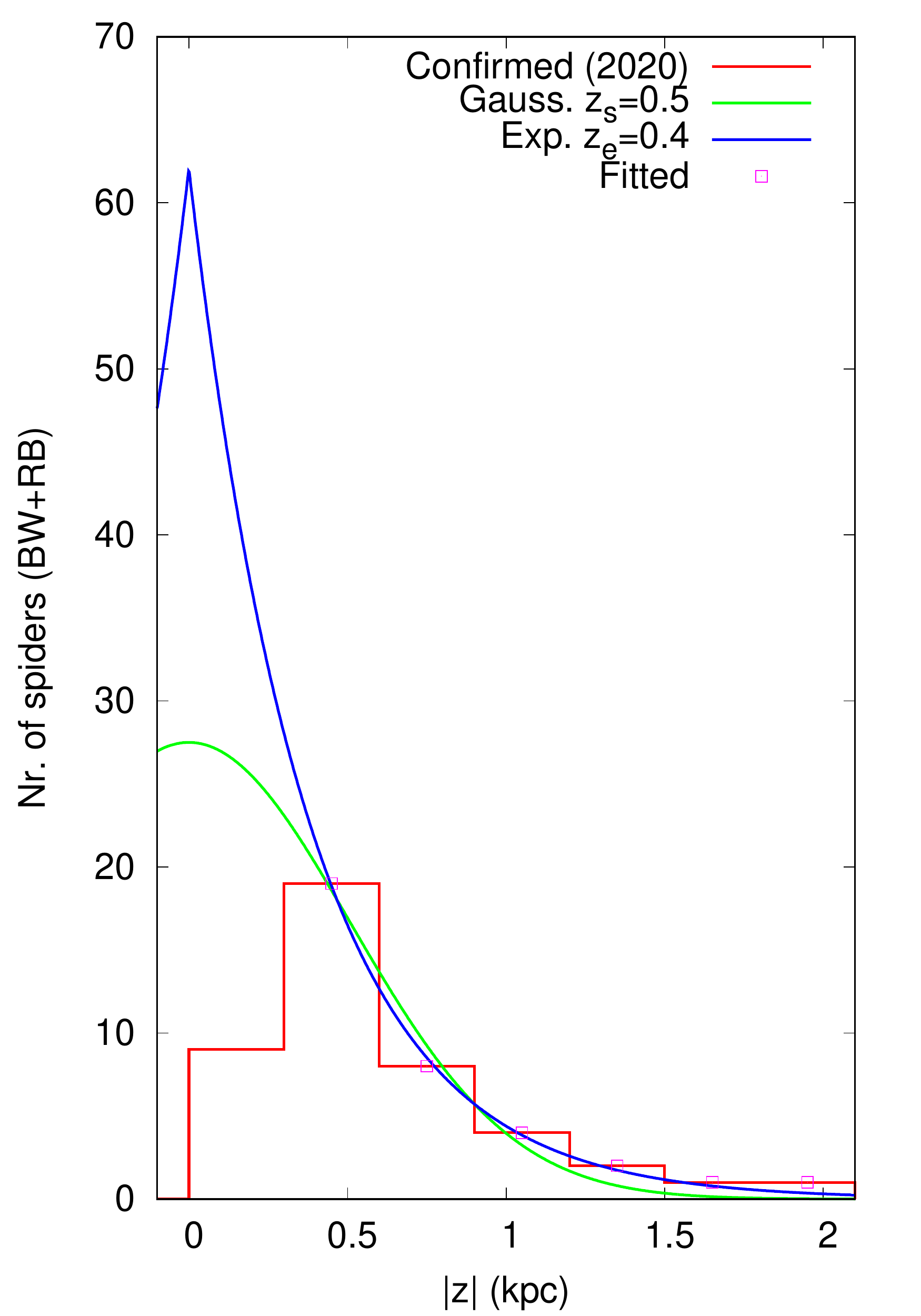}
\caption{\label{fig:xyz} {\it Left:} Top view of the $x$ and $y$
  coordinates of the 52~compact binary MSPs studied in this work. Red
  circles and black squares show redbacks and black widows,
  respectively (open symbols show candidate pulsars). {\it Right:}
  Histogram (red line) of the number of confirmed CBMSPs in the  Galactic
  field as function of the absolute value of the height $z$ above/below the
  Galactic plane. Lines show
  Gaussian (green) and exponential (blue) fits to the data with
  $|z|>0.45$\,kpc.}
\end{figure}

We show in Figure~\ref{fig:xyz} (left) the Galactic Cartesian coordinates of
the 52~currently known CBMSPs; the coordinate system is centered on the Sun
with its $x$ axis pointing towards the Galactic anticenter.
The vast majority of CBMSPs have been discovered in systematic
radio-timing~\cite{Hessels11,Roberts11,Ray12,Bates11,Keith12,Barr13,Camilo15}
or optical-photometric~\cite{Romani12,Kong12,Linares17} searches of
unidentified Fermi-LAT gamma-ray sources.
To avoid diffuse GeV emission near the Galactic plane, most searches
have excluded Galactic latitudes $|\ell| < 5^\circ$.
Thus, there is a strong selection effect against finding CBMSPs near
the plane, which is clearly visible in the $z$ distribution shown
in Figure~\ref{fig:xyz} (right).
Even though the {\it intrinsic\/} number density is expected to increase
when approaching the Galactic plane, the {\it observed\/} density of
CBMSPs drops off sharply below $|z| \simeq 0.5$\,kpc.
From this we can predict
that at least $\sim$20-50 CBMSPs are ``hidden'' in the Galactic plane, with
$|z| \lesssim 0.5$\,kpc.
Those systems should be at distances similar to the currently known population
(less than 6\,kpc) and 
thus we expect them to be detectable as radio pulsars (at higher dispersion measures), 
even though their GeV emission may be blended with the Galactic background. 

Since the current ``horizon" for detection is at approximately 5\,kpc
(all but one CBMSPs in Table~\ref{table:spiders} are within 5\,kpc from
Earth), we estimate that the distribution of $z$ is mostly unbiased
for $|z| > 5\,{\rm kpc}\sin(5^\circ)\simeq 0.45$\,kpc.
By fitting this distribution with an exponential model, we find a
scale height $z_\e=(0.4\pm 0.1)$\,kpc, where the uncertainty is estimated
by using different bin sizes (0.1--0.3\,kpc) and thresholds for the
lowest fitted $z$ values (0.4--0.5\,kpc).
A Gaussian fit gives a slightly higher yet consistent scale height
(0.5\,kpc, see Figure~\ref{fig:xyz}).
This is, to our knowledge, the first determination of the scale height
of CBMSPs.
Previous population synthesis studies including all types of MSPs found
similar values for $z_\e$, $0.5^{+0.19}_{-0.13}$\,kpc~\cite{Cordes97,Story07}.
The population of Galactic low-mass X-ray binaries, from which MPSs
are supposed to evolve, also shows a relatively high scale height
$z_\e=(0.4\pm 0.1)$\,kpc \cite{Grimm02}, in full agreement with our
results for CBMSPs.

\subsection{Total Galactic population}
\label{sec:spiders-total}

In order to estimate the total electron and positron flux from CBMSPs expected
at Earth, we model and simulate the intrinsic Galactic population.
We assume that their number density in the Galactic disk decays
exponentially with the height $|z|$ above the disk and with the distance $R$
to the Galactic center~\cite{Paczynski90,Story07}, 
\begin{equation}
\label{eq:nspi}
n(R,z) = n_\mathrm{c} \,\e^{-R/R_\e} \e^{-|z|/z_\e} \,.
\end{equation}
In order to estimate the local density $n_\odot$ of CBMSPs in the Solar
vicinity, we ignore the $R$ dependence and assume
that the local density of CBMSPs depends only on $z$,
\begin{equation}
\label{eq:nz}
n(z) \simeq n_\odot\, \e^{-|z|/z_\e},  \quad\text{for } R \simeq R_\odot \,.
\end{equation}
We then assume that all systems within the distance
$d_\mathrm{c} \simeq 1$\,kpc from the Sun have already been discovered.
Integrating $n(z)$ within a
cylinder centered around the Sun with radius $d_\mathrm{c}$ and semi-height
$d_\mathrm{c}$ (which approximates the volume where our sample is complete)
yields
\begin{equation}
\label{eq:Ntot}
 N_\mathrm{d\leq d_\mathrm{c}} = 2\pi d^2_\mathrm{c}\, n_\odot z_\e
 \left(1-\e^{-d_\mathrm{c}/z_\e} \right) .
\end{equation}
From the known population of CBMSPs reported in Table~\ref{table:spiders},
we find that the confirmed total number of CBMSPs in the Galactic field
within $d_\mathrm{c} = 1$\,kpc is $N_\mathrm{d\leq d_\mathrm{c}}=13-16$.
From this, using $z_\e=(0.3-0.5)$\,kpc, we estimate the  local density of
CBMSPs as
\begin{equation}
  \label{eq:nsun}
  n_\odot = \frac{N_\mathrm{d\leq d_\mathrm{c}}}{2\pi d^2_\mathrm{c} \,z_\e (1-\e^{-d_\mathrm{c}/z_\e})} \simeq (5-9)~\text{kpc}^{-3} .
\end{equation}
For comparison, Ref.~\cite{Cordes97} found a local density
$29^{+17}_{-11}$\,kpc$^{-3}$ when considering all kinds of MSPs.
In this single exponential approximation, the column density of CBMSP
at the Sun can be determined by integrating Eq.~(\ref{eq:nz}) over all
values of $z$ as $\sigma_\odot=2n_\odot z_\e =(3-9)$\,kpc$^{-2}$.

Now we can use this estimate of $n_\odot$ to normalize the Galactic
density of CBMSPs, assuming a Galactocentric distance for the Sun
$R_\odot=8$\,kpc and neglecting its height above the disk (since it is
$z_\odot \ll z_\e$),
\begin{equation}
\label{eq:n}
n_\mathrm{c} = n_\odot \e^{-R_\odot/R_\e} \simeq (30-60)\,\text{kpc}^{-3} ,
\end{equation}
where we use a radial scale length $R_\e=4.2$\,kpc as found in
Ref.~\cite{Story07}.
Given this value of $R_\e$, $n(R)$ is expected to change by less
than $\sim 25\%$ within $R=R_\odot \pm1$\,kpc, justifying our previous
assumption of a negligible $R$ dependence when estimating $n_\odot$.
Note that all estimates of the density of CBMSPs presented in this
Section ignore beaming effects for the radio pulsar, and thus can be
seen as conservative.
Integrating the Galactic density of CBMSPs within a cylinder of radius 
$d_\mathrm{c}=1$\,kpc for Galactic latitudes between $\ell=-5^\circ$ and
$5^\circ$,
we estimate that there  are 2--3 nearby CBMSPs close to the Sun ($d<1$\,kpc)
and the Galactic plane ($|\ell| < 5^\circ$) which have not 
been discovered yet.
These ``nearby spiders close to the plane" are particularly relevant
for the CR positron flux on Earth, and are discussed further in
Sections~\ref{sec:results} and \ref{sec:summary}.

The extrapolation of these densities far from the Solar vicinity
($R \ll R_\odot$ or $R \gg R_\odot$) is highly uncertain, as there are
currently no CBMSPs detected near the Galactic bulge.
This is not critical for our purpose of studying cosmic ray positrons
with energies well above 20\,GeV, since energy losses make systems at
large distances less important.
With the previous caveat in mind, we can integrate the number density
for $R \rightarrow \infty$ to estimate the total number of CBMSPs in the
Galaxy as
\begin{equation}
\label{eq:Ntot}
N_\mathrm{tot} = 2 \pi n_c R^2_\e \times 2 z_\e  \sim (2-7)\times 10^3 \,.
\end{equation}
In order to estimate the contribution to the observed positron flux of 
CBMSPs that have not been discovered yet, we add a simulated population of
5000~CBMSPs.
More precisely, we choose for a given source its Galactic longitude
$b$ isotropically, while we select $R$ and $z$ according to the
distribution~(\ref{eq:nspi}). Then we determine the source distance to
the Sun projected on the Galactic plane, $d_\mathrm{p}$, and associate
to the source a weight $w$, given by the incompleteness of the known
sources in the torus with the distance $d_\mathrm{p}$. Thus the weight
of the source will vary between $w=0$ for $d_\mathrm{p}<1$\,kpc and
$w=1$ for $d_\mathrm{p}>5$\,kpc. We assume that the known sample of
CBMSPs is unbiased, and approximate the spectral properties of the
simulated population by repeating the values of the 52~known spiders.

\section{Pair spectra from intrabinary shocks}
\label{sec:pairs}

\subsection{A simple model}
\label{sec:pairs-model}

\begin{figure}[tbp]
\centering 
\includegraphics[width=1.0\textwidth,angle=0]{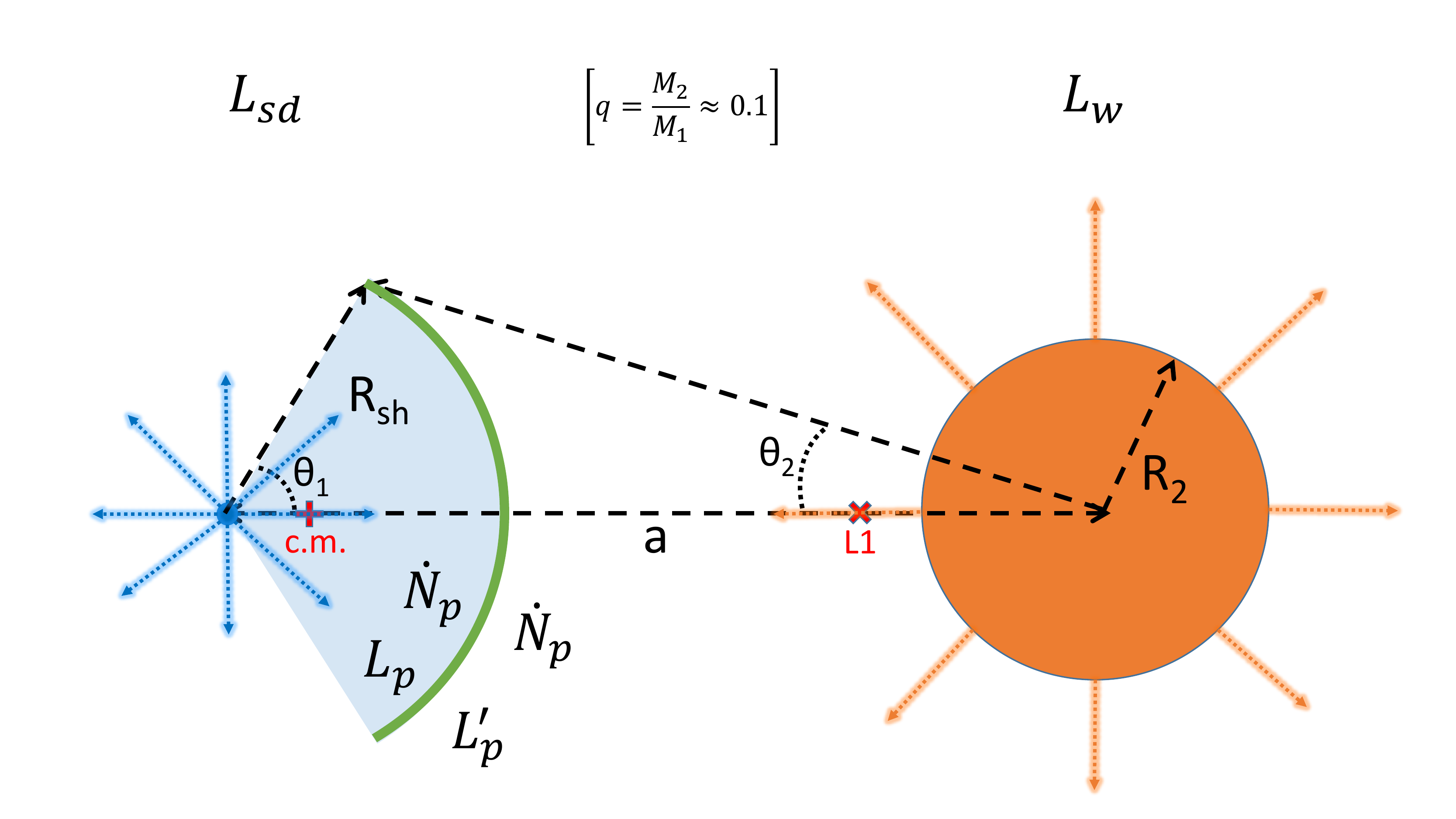}
\caption{\label{fig:model} Sketch of the model for reaccelerated
  $e^\pm$ pairs from intrabinary shocks in CBMSPs. The location of the
  center of mass (c.m.) and inner Lagrangian point (L1) correspond to
  a mass ratio $q=0.1$.}
\end{figure}

We develop a simple model of the intrabinary shock in CBMSPs (sketched
in Figure~\ref{fig:model}) to calculate analytically the spectra and fluxes 
of pairs reaccelerated at the shock.
The shock is assumed to be a spherical shell or cap centered around
the pulsar with opening angle $\theta_1$, so that the fraction of the
pulsar sky covered by the shock is $\Omega_1=(1-\cos\theta_1)/2$.
Indeed, the observed X-ray orbital modulation suggests that in most
CBMSPs the shock is curved around the pulsar (with just one exception
out of nine systems in Ref.~\cite{Wadiasingh17}).
Our "pulsar sky fraction" parameter $\Omega_1$ measures what fraction
of pairs is intercepted by the shock and can thus be reaccelerated.
Likewise, we define $\Omega_2=(1-\cos\theta_2)/2$ as the fractional
solid angle subtended by the shock seen from the companion star, where
$\theta_2$ is the opening angle from the center of the companion and
is related with $\theta_1$ via the orbital separation $a$ and the shock
radius $R_\mathrm{sh}$, cf.\ with Fig.~\ref{fig:model}.

{\it Shock radius}.
We estimate the shock radius $R_\mathrm{sh}$ by
balancing the ram pressures of the pulsar and companion winds, following
the treatement of case~b in Ref.~\cite{Harding90}.
For $R_\mathrm{sh}\ll a-R_2$, where $R_2$ is the companion radius, this leads to
\begin{equation}
\label{eq:Rshock:1}  
\frac{R_\mathrm{sh}}{a} \simeq \frac{\sqrt{A_w}}{1+\sqrt{A_w}}\,,
\end{equation}
where
\begin{equation}
\label{eq:Rshock:2}
A_w = \frac{L_\mathrm{sd}}{\dot{M}_w v_w c}
\end{equation}
is the ratio between pulsar and companion wind ram pressure, with
$\dot{M}_w$ and $v_w$ as the mass loss rate and velocity of the
companion wind, respectively.
%


{\it Companion wind}. We approximate $v_w$ as the escape velocity from
the companion, $v_w$=$\sqrt{2GM_2/R_2}$, where $M_2$ and $R_2$ are the
companion mass and radius, respectively.
We assume that the companion fills its Roche lobe, with a
volume-equivalent radius $R_\mathrm{L2}$.
To estimate the mass loss rate in the companion wind, we use an
"evaporative wind" driven by irradiation from the pulsar wind,
following the formalism of Ref.~\cite{Stevens92}.
In this scenario, a fraction $f_w$ of the irradiating luminosity
intercepted by the companion goes into launching a thermal wind, with
a kinetic luminosity
\begin{equation}
\label{eq:Lw}
 L_\mathrm{w} = \frac{\dot{M}v_w^2}{2}
 =f_w L_\mathrm{sd} \left(\frac{R_2}{2a}\right)^2.
\end{equation}
The mass loss rate in the companion wind then becomes
\begin{equation}
\label{eq:Mdot}
\dot{M}_w = f_w \left(\frac{R_2}{4GM_2}\right) \left(\frac{R_2}{a}\right)^2\,.
\end{equation}


{\it Shock magnetic field and maximal pair energy}.
We then calculate the strength of the pulsar's magnetic field at the
shock, $B_\mathrm{sh}$.
We assume that the field is dipolar inside the light cylinder
($\propto 1/r^3$) and toroidal outside ($\propto 1/r$)~\cite{Harding90}.
The synchrotron loss $t_s$ and diffusive acceleration $t_a$ time scales
at the shock depend on $B_\mathrm{sh}$ and the energy $E$ of electrons and
positrons as~\cite{Harding90},
\begin{equation}
\label{eq:tshock:1}  
t_s = 200 \,\text{s} \left(\frac{\mathrm{G}}{B_\mathrm{sh}}\right)^2 \frac{\mathrm{TeV}}{E},
\end{equation}
and
\begin{equation}
\label{eq:tshock:2}
t_a \simeq \frac{E}{ceB_\mathrm{sh}} \frac{\xi(\xi+1)}{(\xi-1)} = 0.043~\text{s}~\frac{\xi(\xi+1)}{(\xi-1)}~\frac{\mathrm{G}}{B_\mathrm{sh}}~\frac{E}{\mathrm{TeV}} \,,
\end{equation}
where the shock compression ratio $\xi$ is $\simeq 4$ for a strong shock.
By equating these two time scales, we estimate the maximal or cut-off energy
of the pair spectra due to synchrotron losses,
\begin{equation}
\label{eq:Ecut}
 E_\mathrm{cut} =
 68~\text{TeV}~\left(\frac{B_\mathrm{sh}}{\text G}\right)^{-1/2}~
 \sqrt{\frac{(\xi-1)}{\xi(\xi+1)}} \,.
\end{equation}
As noted in Ref.~\cite{Harding90}, $B_\mathrm{sh}$ is high so intrabinary shocks
in CBMSPs are potentially efficient particle accelerators.
Synchrotron losses, however, impose a limit on the maximum energy of
electron and positrons and thus on the kinetic power of the outgoing pairs.
The closer the shock is to the pulsar, the higher is $B_\mathrm{sh}$ and the
lower is $E_\mathrm{cut}$.
For our systems, this limit is in the range
$E_\mathrm{cut}=(1-10)$\,TeV, and the relevant magnetic fields are
$\gtrsim 10$\,G, as we will show in Section~\ref{sec:pairs-results}.
The Larmor radius of the accelerated $e^\pm$ is
$R_\mathrm{L} \simeq 3\times 10^9 \text{cm}~(E/\mathrm{TeV}) (\mathrm{G}/B)
\lesssim 10^9\,\text{cm}$, which is in any case more than ten times
smaller than the size of the accelerating region ($R_\mathrm{sh}
\gtrsim 6 \times 10^{10}$\,cm).
Therefore, the pairs can be accelerated to the relevant energies
before escaping the shock and the synchrotron limit prevails.


{\it Input pair spectra}. The input or ``secondary"
pairs\footnote{``Secondary" meaning after pair cascades in the
  magnetosphere, see Section~\ref{sec:intro}.} entering the shock are
assumed to have a minimum energy $E_{\min}$ between 1 and 50\,GeV.
This range is taken from the pair cascade simulations of Harding and
Muslimov tailored to MSPs~\cite{Harding11}, cf.\ with their Figure~10.
The pair spectra they find are relatively soft, $\d N/\d E\propto E^{-\alpha}$
with $\alpha\simeq 3$, with
low- and high-energy cutoffs at $E_{\min}=1-50$\,GeV and $\sim 1$\,TeV,
respectively.
Because we are mainly interested on the cosmic-ray positron excess
seen at Earth well above 20\,GeV, we only use this ``secondary" component as
input for shock acceleration, and neglect the direct contribution to
the positron flux at Earth. This direct contribution was deemed
negligible by Venter {\it et al.\/} when
studying all Galactic MSPs~\cite{Venter15}.

The total rate of pairs emitted from each polar cap, and
the luminosity that these carry, depend mostly on the spin-down
luminosity of the pulsar.
Using the results of Ref.~\cite{Harding11}, from
their intermediate case with offset polar cap parameter
$\epsilon=0.2$, the total ``incoming" pair rate that reaches the
shock is
\begin{equation}
\label{eq:Ndot}
\dot{N}_\mathrm{p} = 2\Omega_1 \times 8.5\times10^{33}\,\text{s}^{-1}\left(\frac{L_\mathrm{sd}}{10^{35}\,\text{erg\,s}^{-1}}\right)^{0.91} \,,
\end{equation}
and the total "incoming" pair luminosity that reaches the shock is
\begin{equation}
\label{eq:Lpair}
L_\mathrm{p} = 2\Omega_1 \times 3.2\times10^{32}\,\text{erg\,s}^{-1}\left(\frac{L_\mathrm{sd}}{10^{35}\,\text{erg\,s}^{-1}}\right)^{0.86} \,,
\end{equation}
both of which we use below to normalize the post-shock or "tertiary"
pair spectrum.


{\it Shock acceleration}. At the shock, pairs can gain energy via the
so-called Fermi first-order acceleration when travelling back and
forth across the termination front~\cite{Drury:1983zz}.
Such ``reacceleration" will harden the pair spectrum as
compared to the secondary/incoming one, and increase the maximum
energy of the tertiary/outgoing pairs.
In particular, first-order Fermi acceleration in strong non-relativistic
shocks results in a universal spectral energy distribution with power-law
index $\simeq 2$~\cite{Drury:1983zz}.
Although the shock is trans-relativistic, we will assume that a power-law
energy distribution with index $\simeq 2$ is a reasonable approximation to
the post-shock electron-positron spectrum.

{\it Normalization and energy balance}. Using these assumptions, the
differential rate of shock accelerated pairs between $E_{\min}$ and $E_{\max}$
is given by
\begin{equation}
\label{eq:pairspec}
Q(E) = \frac{\d N}{\d E\,\d t } = K E^{-2}  \,,
\end{equation}
where $K$ is a normalization constant. 
This simple analytical model allows us to calculate the integrated
tertiary pair rate,
\begin{equation}
\label{eq:intspec:1}  
\dot{N}'_\mathrm{p} = \int_{E_{\min}}^{E_{\max}} K E^{-2} \d E = K \left(\frac{1}{E_{\min}} - \frac{1}{E_{\max}}\right) \simeq \frac{K}{E_{\min}} ,
\end{equation}
where the last step is valid for $E_{\max} \gg E_{\min}$,
and the integrated tertiary pair luminosity,
\begin{equation}
\label{eq:intspec:2}
L'_\mathrm{p} = \int_{E_{\min}}^{E_{\max}} K E^{-1} \d E
             = K \ln{\frac{E_{\max}}{E_{\min}}}  .
\end{equation}


To normalize the post-shock/tertiary pair
spectra, we assume that the number of pairs is conserved,
\begin{equation}
\label{eq:norm:1}  
\dot{N}_\mathrm{p} = \dot{N}'_\mathrm{p} \simeq \frac{K}{E_{\min}}\,,
\end{equation}
and that the maximum integrated pair luminosity is the sum of three
terms: i) the incoming/secondary pair luminosity; ii) an additional
fraction $\sigma$ of the spin-down luminosity intercepted by the
shock, and iii) the kinetic luminosity of the companion's wind which
is intercepted by the shock. That is,
\begin{equation}
\label{eq:norm:2}
L'_\mathrm{p,max}  = L_\mathrm{p} + \sigma \Omega_1 L_\mathrm{sd} + \Omega_2 L_\mathrm{w} \,,
\end{equation}
and the maximum energy allowed by the available luminosity at the
shock, Eq.~(\ref{eq:intspec:2}), becomes
\begin{equation}
\label{eq:Emax}
E_{\max}  = E_{\min} \e^{L'_\mathrm{p,max}/K} \,.
\end{equation}

\begin{figure}[tbp]
\centering
\includegraphics[width=.49\textwidth,angle=0]{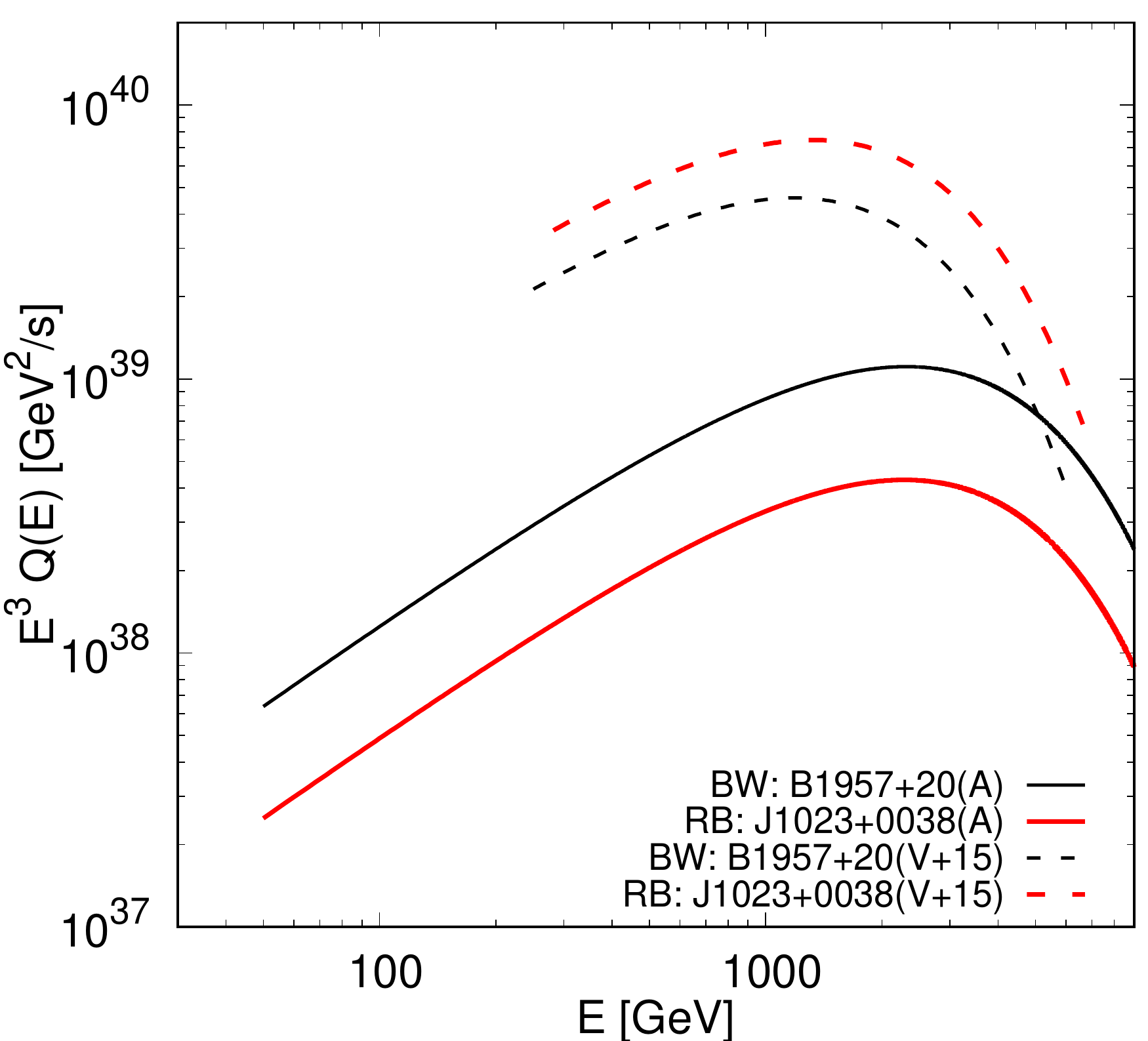}
\includegraphics[width=.49\textwidth,angle=0]{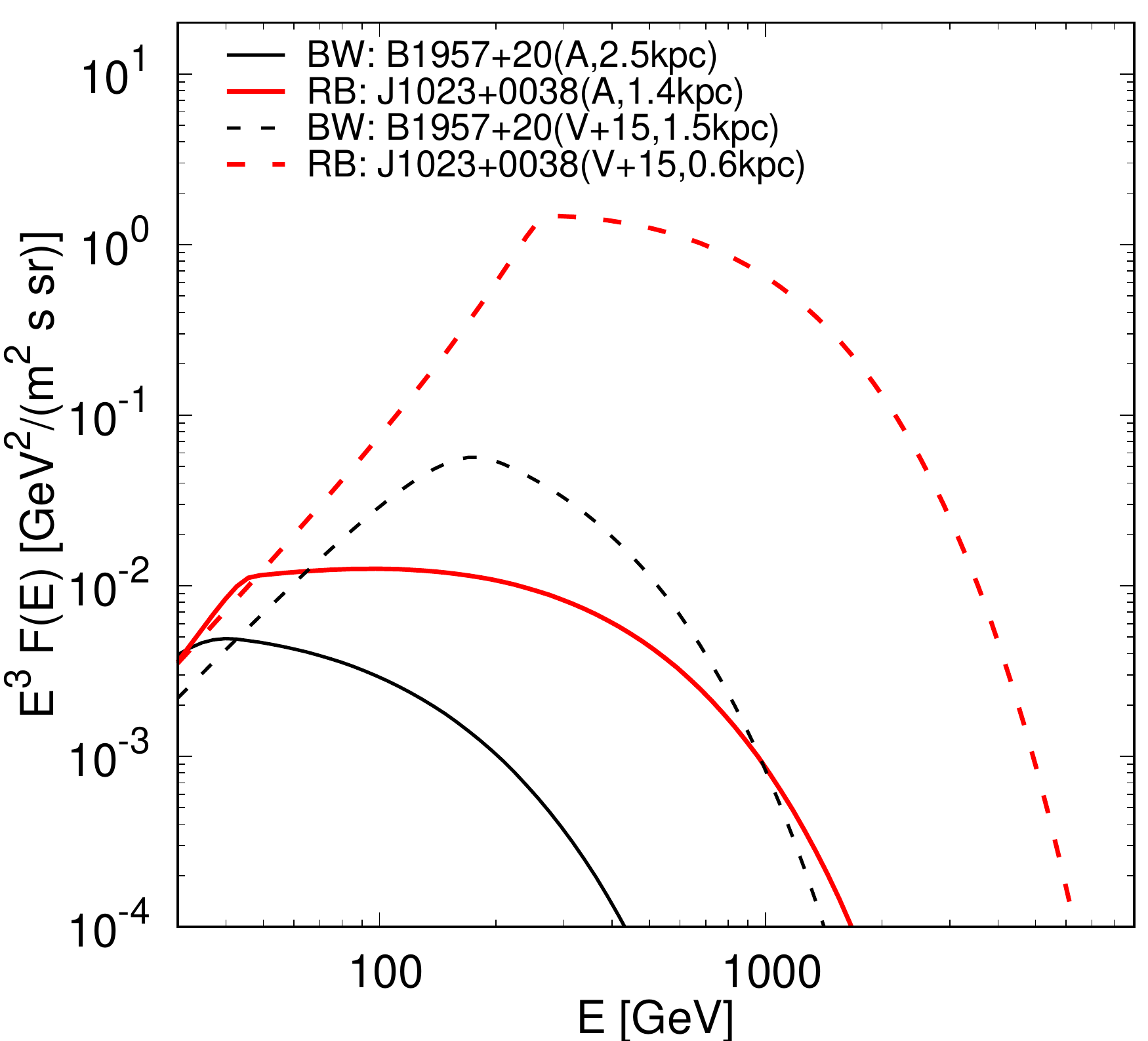}
\caption{\label{fig:compspec} {\it Left:} Injected pair spectra from
  our model A (solid lines) for the prototypical black widow and
  redback systems B1957+20 and J1023+0038 shown in black and red,
  respectively, compared with the injected spectra (dashed lines) from
  Ref.~\cite{Venter15} (their case $\epsilon =
  0.6,\,\eta_\mathrm{p,max}=0.3$). {\it Right:} Diffused pair spectra
  on Earth, for the same two systems. Solid lines show our results
  with updated distance measurements, dashed lines show previous work
  from Ref.~\cite{Venter15}.}
\end{figure}

%
We find the last term in Eq.~(\ref{eq:norm:2}) negligible in most cases,
since $\Omega_2 L_\mathrm{w}$ is more than ten times lower than
$L_\mathrm{p}$.
We see that besides $L_\mathrm{sd}$ and geometric factors ($\Omega_1$ and 
$\Omega_2$), the additional parameter $\sigma$ determines the luminosity of
the outgoing pairs and $E_{\max}$.
We have defined $\sigma$ as the fraction of the intercepted spin-down
luminosity which is available at the shock as kinetic energy to power
Fermi shock acceleration.
Thus $\sigma$ reflects the present uncertainty on the magnetic to
particle energy content in the pulsar wind (the so-called ``sigma
problem" mentioned in Section~\ref{sec:intro}, although note that our
definition does not necessarily follow previous literature).

The lower of $E_{\max}$ and $E_\mathrm{cut}$ gives the true upper
limit on the pair energies, which we call $E_\mathrm{top}$.
Note that we add a high-energy exponential cut-off
$\e^{-E/E_\mathrm{top}}$ to simulate a more physical spectrum.
We can finally find the post-shock (tertiary) pair luminosity,
taking into account synchrotron losses,
\begin{equation}
\label{eq:Lp2}
L'_\mathrm{p} = \int_{E_{\min}}^{\infty} K E^{-1} \e^{-E/E_\mathrm{top}} \d E \,.  
\end{equation}

Summarizing, our model parameters are:
\begin{itemize}
  \item E$_{\min}$: The minimum pair energy after shock reacceleration 
  \item $\Omega_1$: fraction of the pulsar sky covered by the shock (or fraction of emmitted pairs intercepted by the shock)
  \item f$_\mathrm{w}$: fraction of the intercepted pulsar wind that launches the companion's wind
  \item $\sigma$: fraction of intercepted spin-down luminosity available for shock acceleration
\end{itemize}
and our main assumptions:
\begin{itemize}
  \item The post-shock tertiary pair spectrum follows a power law
    with index $-2$.
  \item $E_{\min}$ is the same minimum energy of the secondary pairs
    produced in the pulsar magnetosphere; 
  \item The maximum energy in the post-shock (tertiary) pair spectra
    ($E_\mathrm{top}$) is limited by either the kinetic luminosity available
    at the shock (which can accelerate pairs up to $E_{\max}$) or by
    synchrotron losses (which cut off energies above $E_\mathrm{cut}$,
    Eq.~(\ref{eq:Ecut})), i.e., $E_\mathrm{top}=\min(E_{\max},E_\mathrm{cut})$.
\end{itemize}

We show in the left panel of Figure~\ref{fig:compspec} the spectra injected
into the interstellar medium  from our model, for the prototypical black
widow and redback systems B1957+20 and J1023+0038. In addition, previous
results from Ref.~\cite{Venter15} are shown by dashed lines. This comparison
shows how the large values of $E_{\min}$ assumed by these authors  increase
the normalisation of $Q(E)$ by more than an order of magnitude relative to
our results. Note that the large values of $E_{\min}$ used in
Ref.~\cite{Venter15}
contradict the basic priniciples of Fermi acceleration:  Since at each shock
crossing there is a non-zero escape probability, the accelerated energy
spectrum has to extend down to the energy with which electrons and
positrons are entering the acceleration region.
We will model the diffusion of the injected pairs to Earth in
Section~\ref{sec:flux}. Here, we show an example for the  resulting fluxes
in the right panel of
Figure~\ref{fig:compspec}. In addition to an overall reduction of the
flux caused by our normalisation condition relative to the results of
Ref.~\cite{Venter15}, the increased distances of the two sources lead to
a suppression of the high-energy tail.

\subsection{Model results}
\label{sec:pairs-results}

We explore three different models (defined by our four parameters as
summarized in Table~\ref{table:models}) in order to cover the most
likely configuration of the intrabinary shock and the injected pair
spectra.
These are referred to as models A, B and C, in order of decreasing pair
rate and luminosity.
We refer to the diffuse positron fluxes predicted by these models for
the simulated population of CBMSPs as A$''$, B$''$ and C$''$,
respectively.
When possible, we also compare our results with those for the 24 CBMSPs
previously studied in Ref.~\cite{Venter15}, which we label as V15.
Additionally to these samples, we consider also the possibility that
in the future a close-by spider at the distance $d=0.5$\,kpc will be
found. In the case of anisotropic diffusion, the relative position of
the sources to the local magnetic field line matters. We distinguish
therefore two cases: In case D, we assume that the spider is connected
with the Sun by a magnetic field line, $d\simeq d_\|$, while in case E
the spider is situated orthogonal to the local magnetic field line,
$d\simeq d_\perp$. Here, $d_\|$ and $d_\perp$ denote the source
distance projected along and perpendicular to the local magnetic field
line, respectively.

\begin{figure}[ht]
\centering 
\includegraphics[width=.49\textwidth,angle=0]{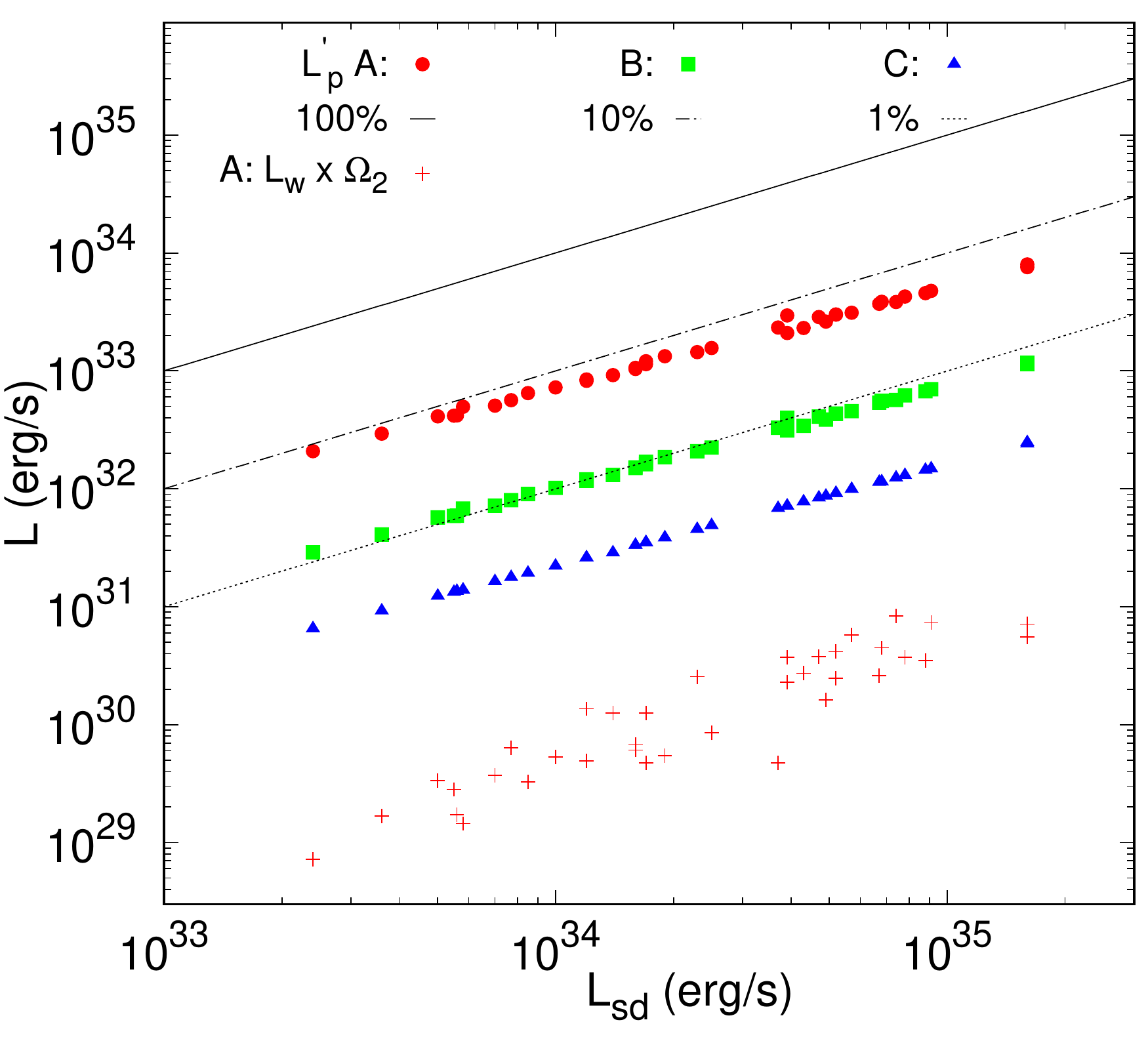}
\includegraphics[width=.49\textwidth,angle=0]{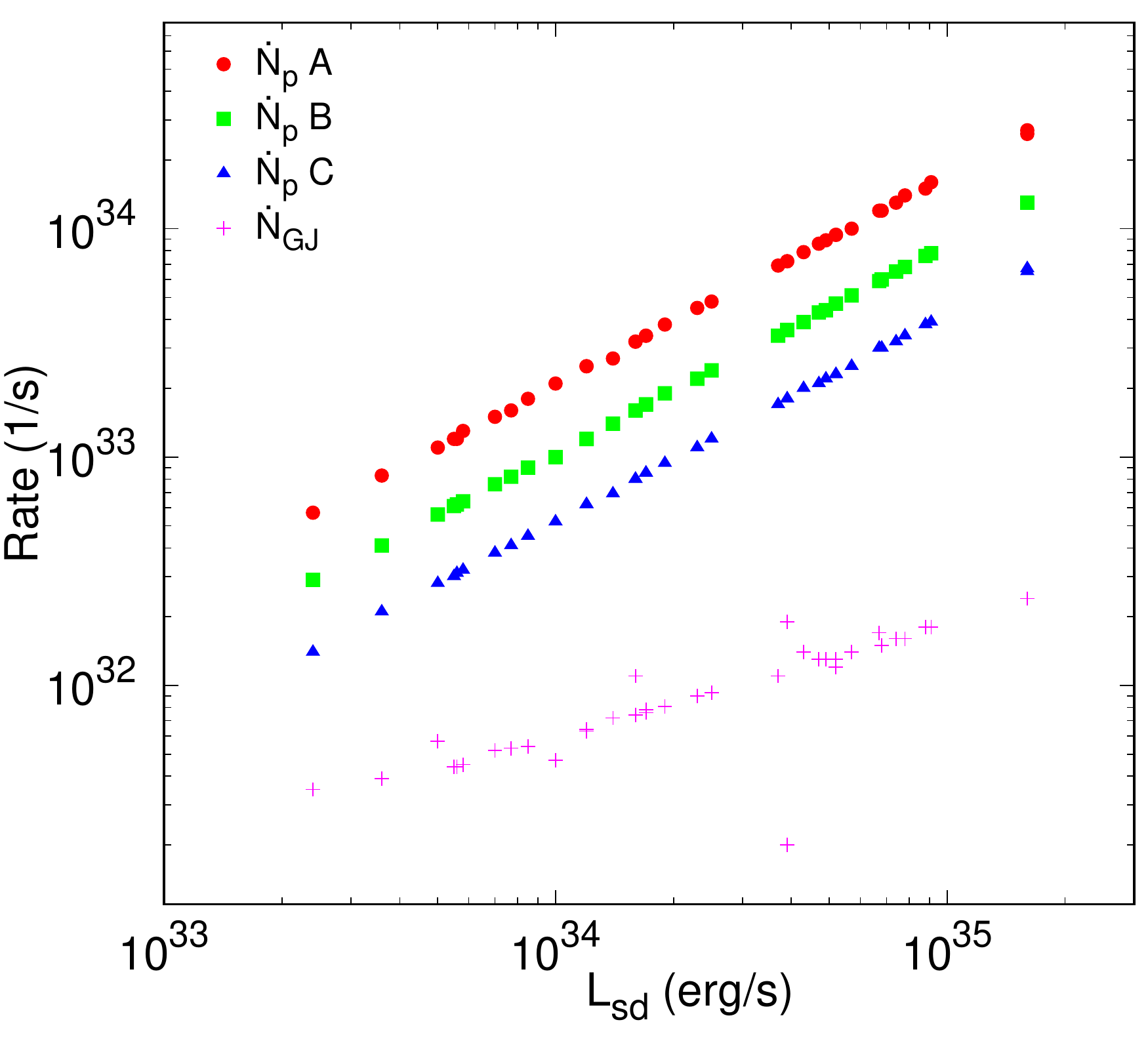}
\caption{\label{fig:ABC} Integrated pair rates (right) and
  luminosities (left) vs. spin-down luminosity for the 52 CBMSPs
  studied in this work, from our models A, B and C (red circles, green
  squares and blue triangles, respectively). Lines on the left panel
  show 1\%, 10\% and 100\% efficiency in converting spin-down
  luminosity into pairs reaccelerated at the shock. The companion
  kinetic luminosity intercepted by the shock (left, red plus signs)
  and the Goldreich-Julian rate (right, magenta plus signs) are also
  shown for comparison.}
\end{figure}

\begin{table}
\centering
\begin{tabular}{|lll|cccc|}
\hline
Label & Nr. & Sample & $E_\mathrm{min}$ & $\Omega_1$ & $f_\mathrm{w}$ & $\sigma$ \\
 &  &  &  (GeV) &  &  &  \\
\hline
V15 & 24 & known in 2015 & 30-1850  & 1.0 & - (0?) & - ? \\
A$'$   & 52     & currently known          & 50 & 1.0 & 0.5    & 0.3 \\
A$''$ & 5000 & simulated outside 1~kpc & 50 & 1.0 & 0.5    & 0.3 \\
B$'$  & 52 & currently known. & 10 & 0.5 & 0.2    & 0.15 \\
B$''$ & 5000 & simulated outside 1~kpc & 10 & 0.5 & 0.2    & 0.15 \\
C$'$   & 52 & currently known. & 4 & 0.25 & 0.1    & 0.0  \\
C$''$ & 5000 & simulated outside 1~kpc & 4 & 0.25 & 0.1    & 0.0  \\
D & 1    & plane perpendicular & 50 & 1.0 & 0.5    & 0.3 \\
E & 1    & plane parallel  & 50 & 1.0 & 0.5    & 0.3 \\
\hline
\end{tabular}
\caption{\label{table:models} Models and samples of CBMSPs explored in
  this work.}
\end{table}

Figure~\ref{fig:ABC} shows the luminosities ($L'_\mathrm{p}$, left) and
tertiary (post-shock) pair rates ($\dot{N}_\mathrm{p}$, right)
for the 52 spiders in our sample, calculated with our models A, B and
C.
The close to linear dependence on $L_\mathrm{sd}$ is readily visible,
inherited from the MSP pair cascade simulations (Eqs.~(\ref{eq:Ndot}) and
(\ref{eq:Lpair}), from \cite{Harding11}).
We find that the tertiary pair rates are between 4 and 100 times
$\dot{N}_\mathrm{GJ}$, while the tertiary pair luminosities are
between 0.1\% and 10\% of the spin-down luminosity $L_\mathrm{sd}$.
We also find that, as expected and as mentioned in
Section~\ref{sec:pairs-model}, the contribution of the kinetic
luminosity from the companion's wind $\Omega_2 L_\mathrm{w}$ is
negligible, between one and three orders of magnitude lower than
$L'_\mathrm{p}$ (Fig.~\ref{fig:ABC}, left).

\begin{figure}[h]
\centering 
\includegraphics[width=.49\textwidth,angle=0]{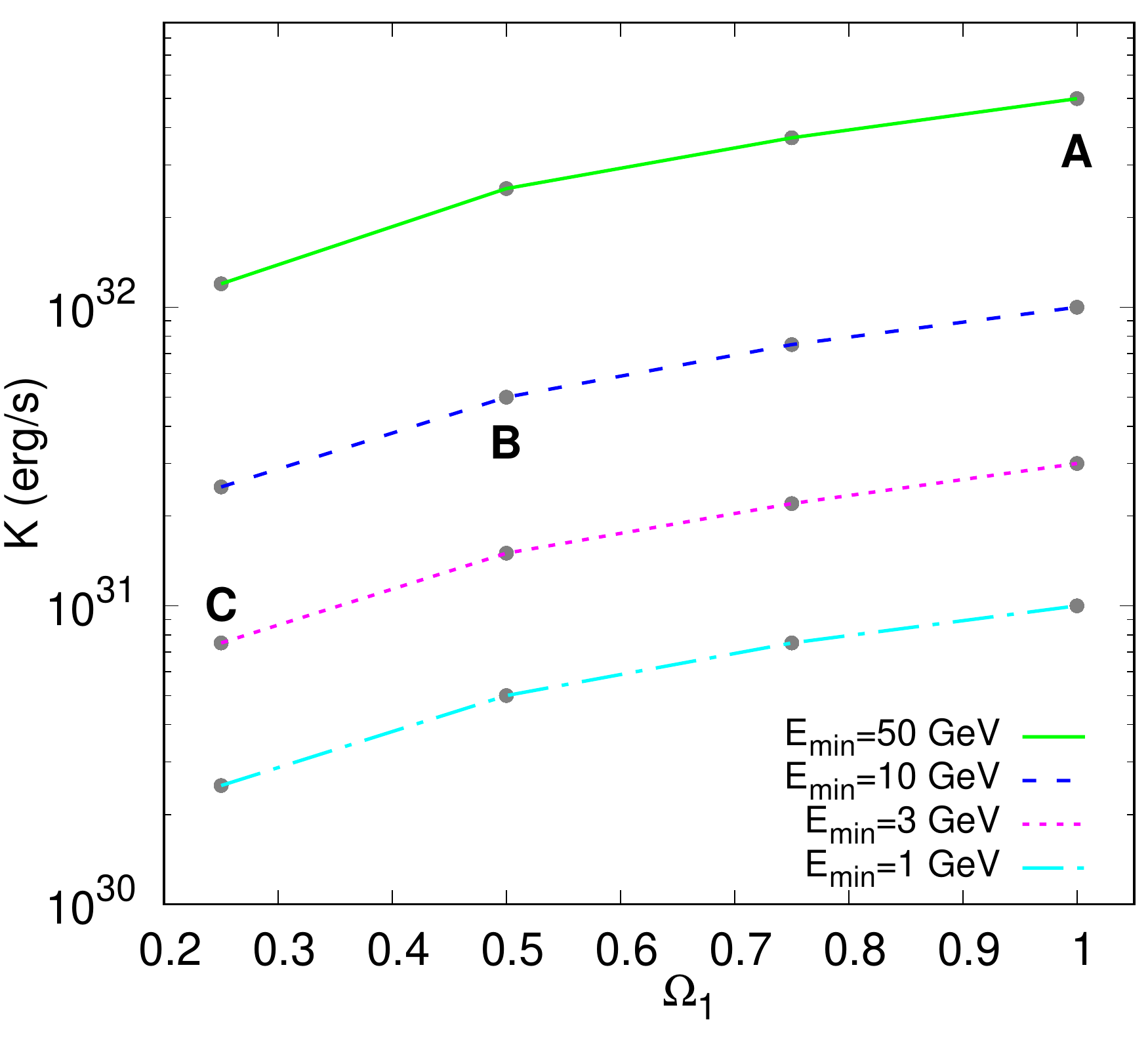}
\includegraphics[width=.49\textwidth,angle=0]{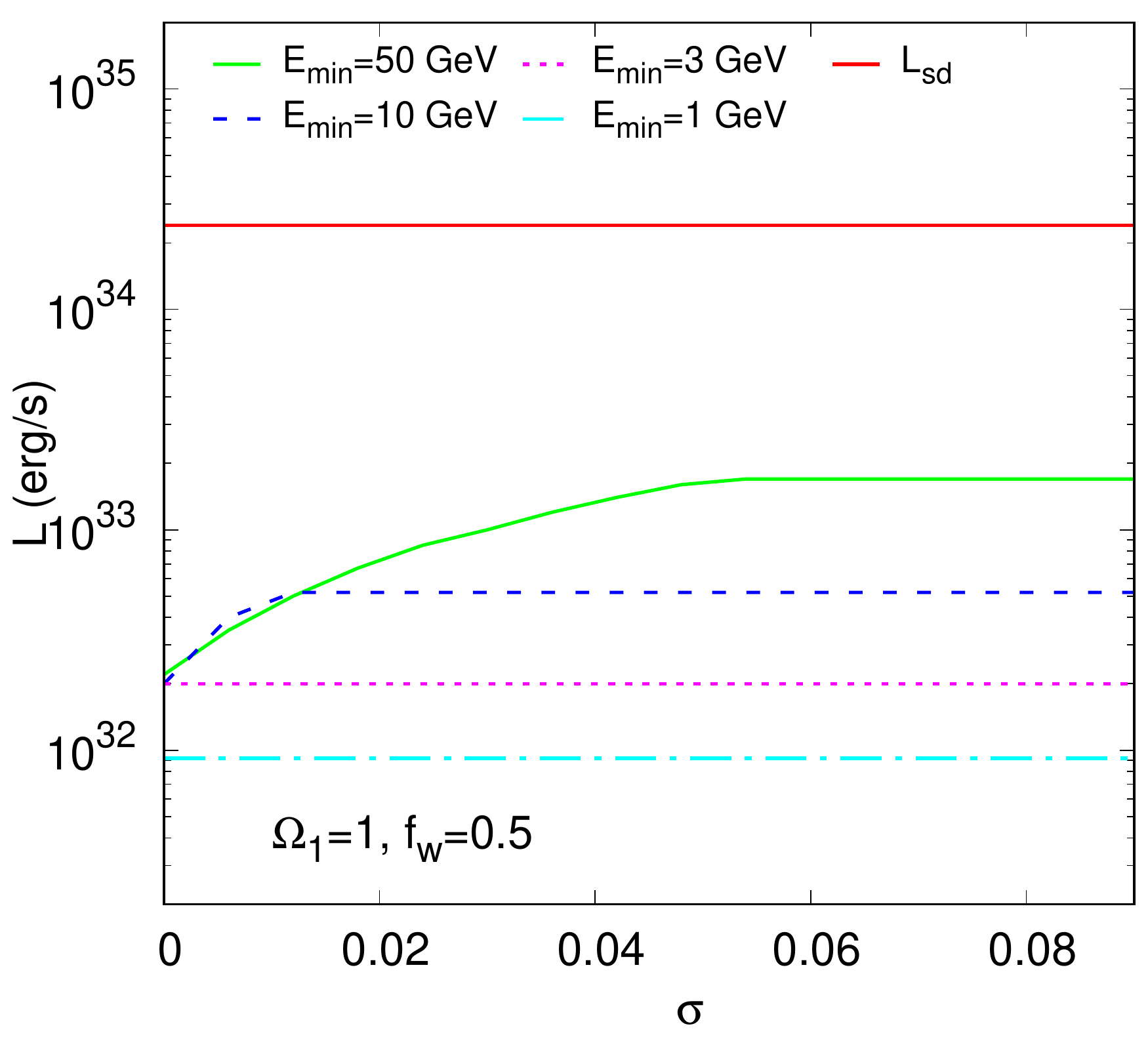}
\caption{\label{fig:KL} {\it Left:} Average normalization of
  the post-shock pair spectra as a function of $\Omega_1$ (fraction of
  the pulsar sky covered by the shock), for different values of
  $E_{\min}$. The location of our ABC models is shown
  with bold letters.  {\it Right:} Average post-shock pair luminosity
  as a function of $\sigma$ (fraction of intercepted $L_\mathrm{sd}$
  that can accelerate pairs), for different values of $E_{\min}$.}
\end{figure}


The most important parameter in setting the tertiary pair rates and
luminosities is the minimum energy of the pairs injected into the
shock, $E_{\min}$.
Since the integrated rate is dominated by the lowest energies
($Q(E)\propto E^{-2}$), the injected rate and minimum energy together
determine the normalization $K$, cf.\ with Eq.~(\ref{eq:norm:1}).
Thus the assumed $E_{\min}$ is the key for a correct normalization of
the injected and observed pair fluxes.
This is shown in Figure~\ref{fig:KL}, where $K$ and $L'_\mathrm{p}$
are plotted as a function of $\Omega_1$ and $\sigma$, respectively,
for different values of $E_{\min}$ (this and the subsequent Figure~6 show
average values in our sample of 52 CBMSPs).
We see that, depending on mostly $E_{\min}$ but also $\Omega_1$, $K$
can vary by a bit more than two orders of magnitude.

In general, the outgoing pair luminosity $L'_\mathrm{p}$ is limited by
synchrotron losses and is at most 10\% of $L_\mathrm{sd}$ (for
$E_{\min} = 50$~GeV, our model A; Figs.~\ref{fig:ABC} and
\ref{fig:KL}).
It is interesting to note that $L'_\mathrm{p}$ saturates at
10\% of $L'_\mathrm{sd}$ for $\sigma \gtrsim 0.05$, as shown in the 
right panel of Figure~\ref{fig:KL}.
Note also that the pairs injected into the shock carry at most 1\% of
$L_\mathrm{sd}$ (cf.\ with Eq.~(\ref{eq:Lpair}) of Ref.~\cite{Harding11}).
Previous work~\cite{Venter15} assumed that this efficiency $\eta =
L'_\mathrm{p}/L_\mathrm{sd}$ can be up to 30\%, and that pairs always
reach the synchrotron cut-off energy $E_\mathrm{cut}$.
This leads to artificially high values for both $K$ and $E_{\min}$, as
can be seen in the left panel of Figure~\ref{fig:compspec}.

\begin{figure}[tbp]
\centering 
\includegraphics[width=.49\textwidth,angle=0]{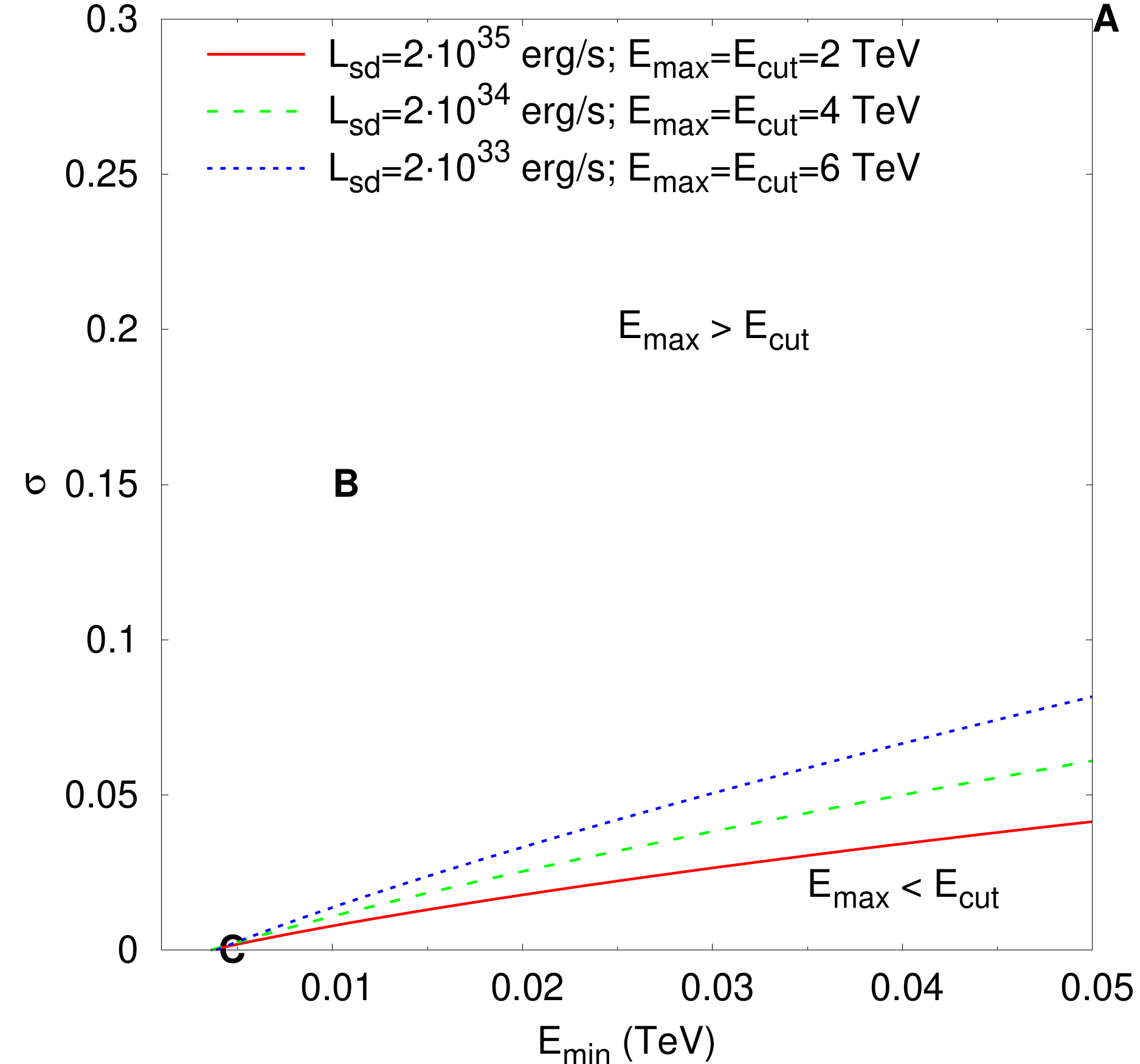}
\includegraphics[width=.49\textwidth,angle=0]{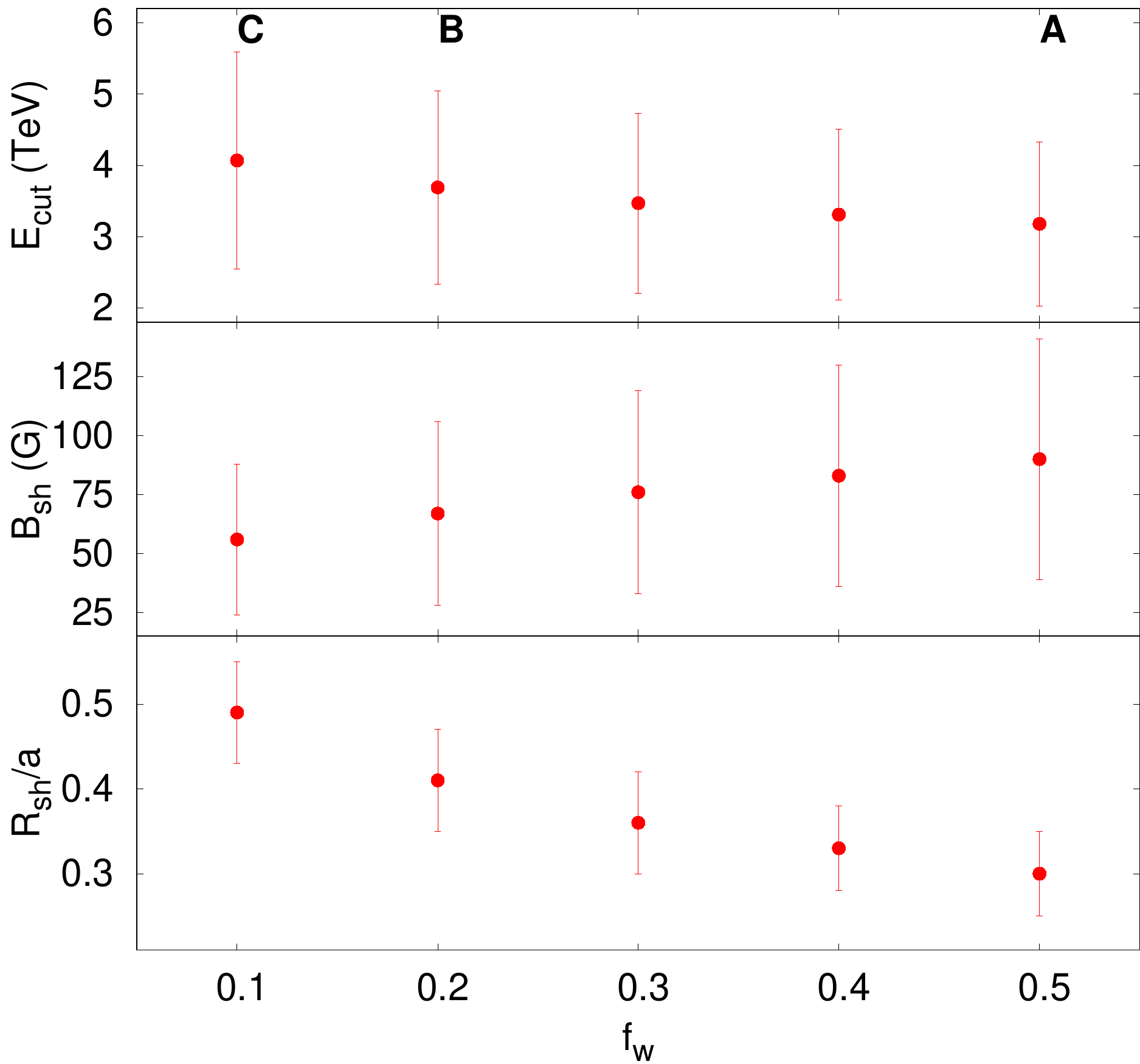}
\caption{\label{fig:Ecut} {\it Left:} Boundaries between the
  synchrotron-limited ($E_{\max} > E_\mathrm{cut} = E_\mathrm{top}$)
  and the power-limited ($E_\mathrm{cut} > E_\mathrm{\max} =
  E_\mathrm{top}$) regimes, for different values of $E_\mathrm{cut}$
  and $L_\mathrm{sd}$ (as indicated).
  {\it Right (from bottom up):} Average and standard deviation of the
  shock radius (in units of the orbital separation, $a$), magnetic
  field and synchrotron cut-off energy as a function of the companion
  wind launching efficiency, $f_\mathrm{w}$.
  In both cases the location of our ABC models is shown with bold letters.
}
\end{figure}

Our model reveals two distinct regimes of intrabinary shock
acceleration, depending on the parameters $E_{\min}$ and $\sigma$, as
shown in the left panel of Figure~\ref{fig:Ecut}.
In what we call the power-limited regime (model C), the maximum pair
energy is not set by synchrotron losses but rather limited by the
available kinetic power at the shock, that is, $E_\mathrm{top} =
E_{\max} < E_\mathrm{cut}$.
This regime occurs only for $E_{\min} \gtrsim 4$~GeV and $\sigma \lesssim
0.05$, with the exact boundaries shown in Figure~\ref{fig:Ecut}.
Instead, in the synchrotron-limited regime (models A and B), there is
abundant kinetic power at the shock to power reacceleration (compared
to $K$, cf. Eq.~\ref{eq:Emax}) so that $E_{\max}$ is high.
This is the more common case where the maximum pair energy is set by
synchrotron losses, i.e., $E_\mathrm{top} = E_\mathrm{cut} <
E_{\max}$.

%
Previous work assumed that all pairs created in the pulsar
magnetosphere can reach the shock and be reaccelerated \cite{Venter15}.
Because this fraction of pairs intercepted by the shock is not tightly
constrained, we conservatively explore in this work the range
$\Omega_1=[0.25-1.0]$, which corresponds to opening angles
$\theta_1=[60^\circ-180^\circ]$. Varying these parameters changes
$\dot{N}_\mathrm{p}$ and $L'_\mathrm{p}$ by a factor
four, since the predicted rates are proportional to $\Omega_1$.
More detailed models of the pressure balance between the pulsar wind
and the companion star magnetosphere/wind show more complex ``bow"
shapes for the intrabinary shock, but a similar range for $\theta_1$
and $\Omega_1$ \cite{Sanchez17,Wadiasingh17,Wadiasingh18}.
%


%
We find that the companion wind has no strong impact on the pair
spectral parameters, with our main assumption of an irradiation driven
wind (Sec.~\ref{sec:pairs-model}).
The range $f_\mathrm{w}=[0.1-0.5]$ that we explore in this work
yields relatively high mass loss rates for the companion, of the order
$10^{15}$--$10^{17}$\,g/s.
We find that the shock radius is anti-correlated with the companion
wind parameter $f_\mathrm{w}$, as shown in Figure~\ref{fig:Ecut}: The
stronger the companion wind is, the closer the shock gets to the
pulsar.
The corresponding magnetic field at the shock, $B_\mathrm{sh}$, is in
the 5--240\,G range for our ABC models (Sec.~\ref{sec:appendix}) and
shows a weak dependence on the companion wind parameter $f_\mathrm{w}$
(Figure~\ref{fig:Ecut}).
These values agree with the available observational constraints on the
magnetic field at the intrabinary shock~\cite{Li19,Polzin19}.
The synchrotron cut-off energy $E_\mathrm{cut}$, which is always in the range
2--10~TeV, also shows a weak dependence on $f_\mathrm{w}$.

\section{Galactic transport with regular and turbulent fields}
\label{sec:flux}

The transport equation for the differential number density $n(\vec x,t,E)$
of positrons including energy losses  $b(E,t)$ is given by
\be \label{diffloss0}
\frac{\partial n}{\partial t} - \vec\nabla_i D_{ij} \vec\nabla_j n
  -\frac{\partial}{\partial E}(b n) =  Q .
\ee
In the presence of a regular, uniform magnetic field, the diffusion
tensor $D_{ij}$ can be written as
\be
 D_{ij} = D_\| e_ie_j + D_\perp(\delta_{ij}-  e_ie_j) \,.
\ee
Here, $\vec e$ is a unit vector in the direction of the regular magnetic
field, while the diagonal elements of the diffusion tensor describe diffusion
along ($D_{\parallel }$) and perpendicular ($D_{\perp }$) to the regular field.
In Ref.~\cite{Giacinti:2017dgt}, these diffusion coefficients were calculated
as function of the ratio of the strength of the regular and the turbulent
field, $\eta=B_{\rm rms}/B_0$. Since  charged particles can move freely along
the direction of the regular field, while they gyrate along the line in the
perpendicular directions, there is a strong ordering of the two diffusion
coefficients, $D_{\parallel }\gg D_\perp$, for $\eta\ll 10$. Comparing then
the resulting grammage crossed by CRs to measurements, the authors of
Ref.~\cite{Giacinti:2017dgt} determined a range of allowed values for the
diffusion parameters. In particular, they argued that in the
Jansen-Farrar model~\cite{Jansson:2012pc} for the regular
Galactic magnetic field, a value of $\eta\approx 0.25$ is consistent with
the average angle of $20^\circ$ between magnetic field lines and the
Galactic plane. The need for anisotropic CR diffusion was previously
stressed in Refs.~\cite{2014ApJ...785..129K,Giacinti:2015hva}.

The solution of Eq.~(\ref{diffloss0}) can be found
either using the Green function method~\cite{1959SvA.....3...22S} or by
Fourier transforming its spatial part~\cite{Berezinsky:2005fa},
introducing the Syrovatskii variables
\be
\lambda_i^2(E,E_g)= \int_{t_g}^{t} \d t^{\prime} \,D_i(E(t^{\prime}))
= \int_E^{E_g} \d E^{\prime} \,\frac{D_i(E^{\prime})}{b(E^\prime)}
\ee
with $i=\{\perp,\|\}$. These new variables correspond to the squared distance
traveled by a particle perpendicular and parallel to the magnetic field line,
while its energy diminishes from $E_g$ to $E$.
Assuming for the energy scaling of the diffusion coefficient a power law,
$D_i=D_{0,i}(E/E_0)^\delta$, and using quadratic energy losses
$b=-\d E/\d t=\beta E^2$ typical for Thomson and synchrotron losses, one can
find a closed expression for the Syrovatskii variable,
\begin{equation}
\label{lambda}
  \lambda_i^2(E,E')  = \frac{D_{0,i}}{(\delta-1)\beta E} \:
  \left( \frac{E}{E_0} \right)^{\delta}
  \left[\left( \frac{E^\prime}{E} \right)^{\delta-1} -1 \right] .
\end{equation}
The case of anisotropic diffusion can be reduced to isotropic diffusion,
recalling that the isotropic Green function factorizes in Cartesian
coordinates. Alternatively, one can first diagonalise $D_{ij}$ by a
rotation, and then perform a scale transformation. In either case, one
finds for the case of a single source at the distance $r$,
\be   \label{diff_cont_E}
n(\vec r) = \int\d E'\,
  \frac{Q(E')}{b(E)} \:
\frac{\exp\left(-\frac{\vec r_\perp^2}{4\lambda_\perp^2}
\right)\exp\left(-\frac{\vec r_\|^2}{4\lambda_\|^2} \right)}
      {\left( 4\pi\right)^{3/2} \lambda_\perp^2\lambda_\|} .
\ee
We calculate $r_\perp$ and $r_\|$ by projecting $\vec r$ on the unit
vector along the magnetic field line going through the source. The
direction of the magnetic field line is determined in turn by using
the spiral field of the Jansen-Farrar model~\cite{Jansson:2012pc} for
its $x$-$y$ components and adding an $X$ field-like component with
angle $\theta=20^\circ$ to the Galactic plane. While this toy model neglects,
e.g., the increase of
$|\theta|$ for increasing $z$, it captures the main differences to
the usually considered case of isotropic diffusion. The latter is
obtained by setting $D_{\perp,0}=D_{\|,0}=D_{0}$ in Eq.~(\ref{lambda}).

Finally, we note that the solution~(\ref{diff_cont_E}) is valid in $\MR^3$.
In order to take into account the finite vertical extension $H$ of the
Galactic CR halo, $H\simeq 5$\,kpc, and the resulting escape of CRs, we
multiply therefore the integrand by the function
$\exp(-\lambda_\|\sin\theta/H)$.
For the case of anisotropic diffusion, we set
$D_{\perp,0}=2\times 10^{26}$\,cm$^2$/s and $D_{\|,0}=5\times 10^{28}$\,cm$^2$/s
at $E_0=5$\,GeV, while we use $D_{0}=3\times 10^{28}$\,cm$^2$/s for
isotropic diffusion~\cite{Giacinti:2017dgt}. Moreover, we assume Kolmogorov
diffusion, $\delta=1/3$, motivated by the B/C measurements of
AMS-02~\cite{Aguilar:2016vqr} and use $\beta=4\times 10^{-17}$\,GeV/s.

\section{Diffuse positron flux from spiders}
\label{sec:results}

\begin{figure}[tp]
\centering
\includegraphics[width=.9\textwidth,angle=0]{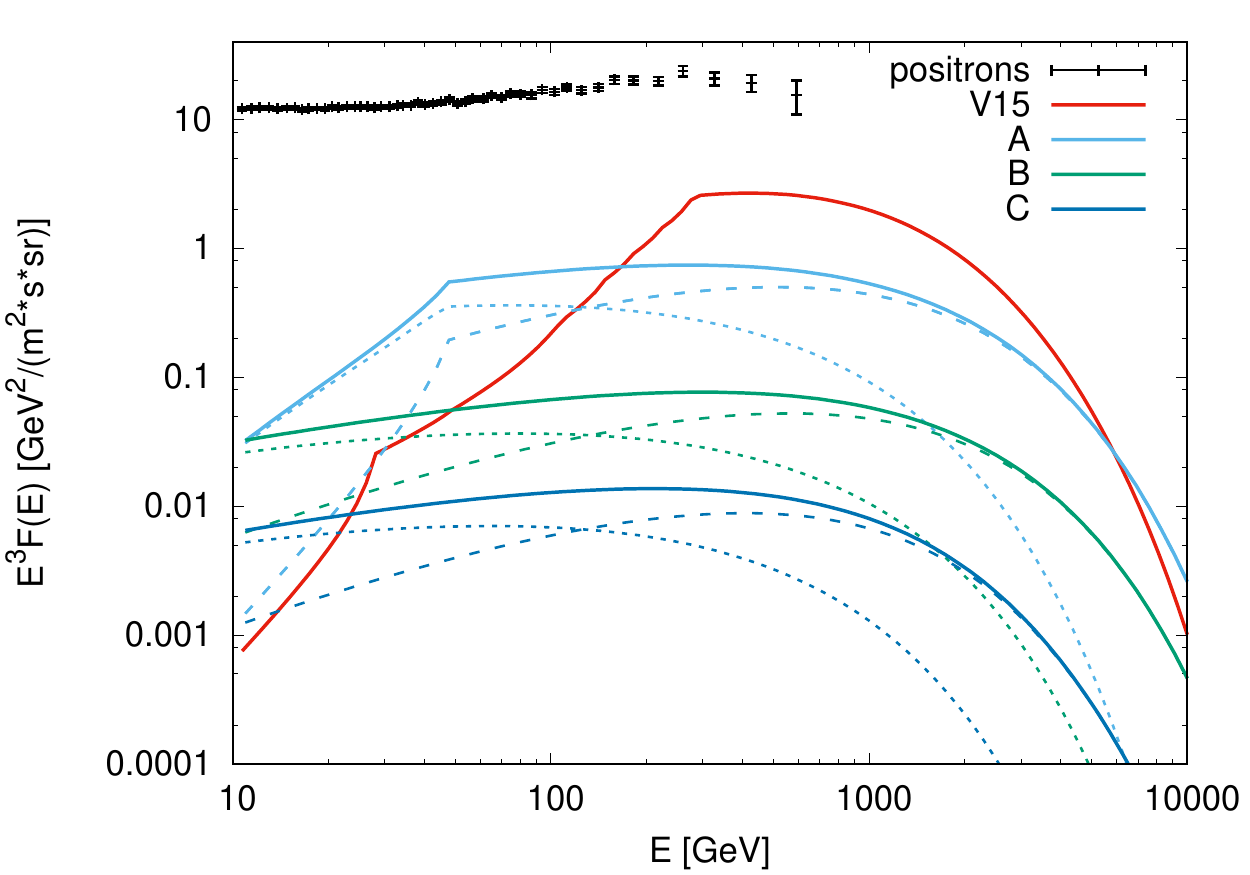}
\caption{\label{fig:Fiso} Positron fluxes on Earth after {\it isotropic} diffusion from our ABC models. Dashed and dotted lines show the contribution from the currently known and simulated population of CBMSPs, respectively, and solid lines show total flux. Black error bars show the AMS-02 data and the red line compares
  to previous work (V15, \cite{Venter15}; see Table~\ref{table:models} and Section~\ref{sec:results}).}
%
\centering
\includegraphics[width=.9\textwidth,angle=0]{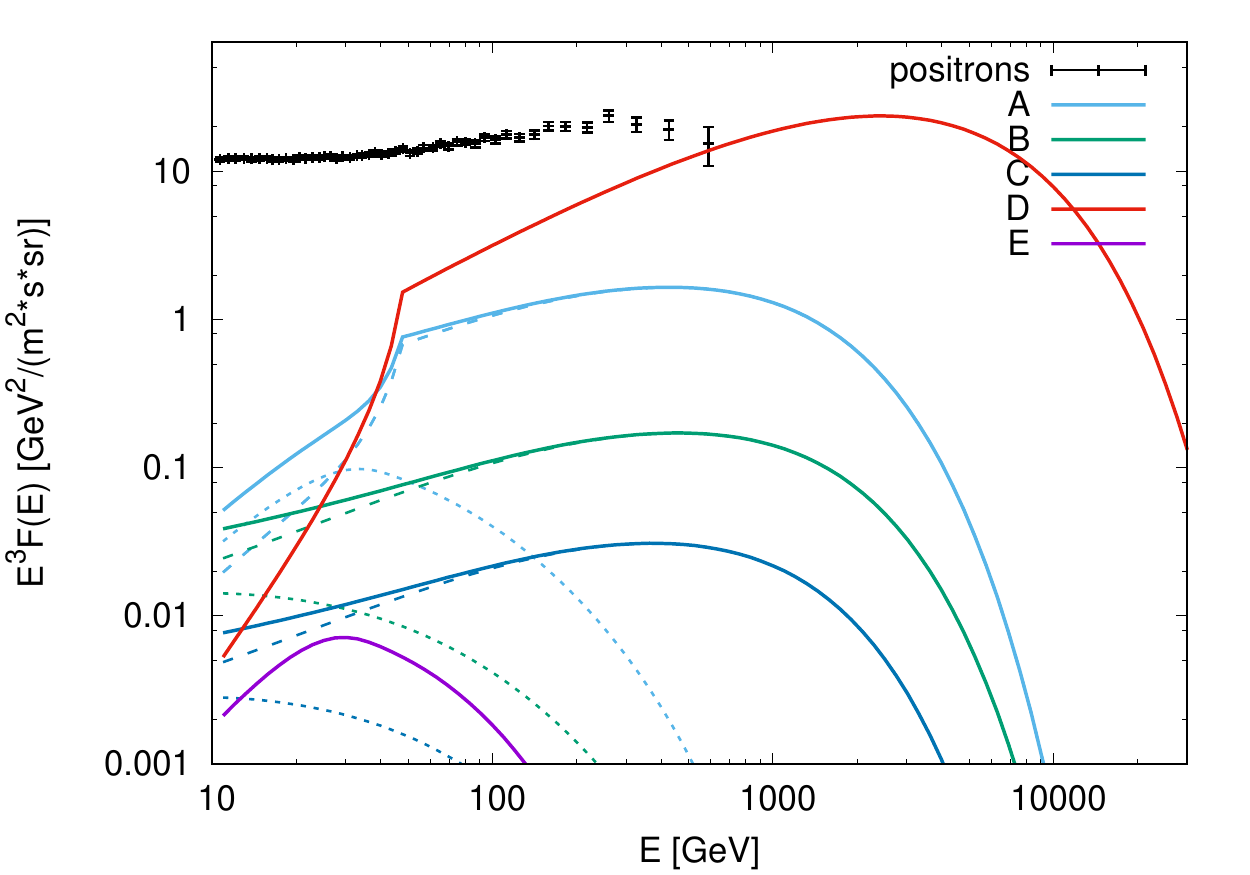}
\caption{\label{fig:Faniso} Positron fluxes on Earth after {\it anisotropic} diffusion from our ABCDE models. Lines labelled D and E show the contribution from one nearby CBMSP on the Galactic plane from a direction which is aligned and misaligned with the regular Galactic magnetic field, respectively (other lines and symbols as in Fig.~\ref{fig:Fiso}; see Table~\ref{table:models} and Section~\ref{sec:results}).}
\end{figure}

As a first step, we compare the diffuse positron flux reported by Venter
{\it  et al.\/} in Ref.~\cite{Venter15} (their Figure~8, cyan lines) with
the diffuse flux that we calculate from their source sample and using
their distances (the 24 spiders known in 2015; red line in
Figure~\ref{fig:Fiso}).
We use for the purpose of this comparison their injected source spectra for
$\epsilon =0.6$ and $\eta_\mathrm{p,max}=0.3$, even if we have argued that
they overestimate both $K$ and $E_{\min}$.
This serves as a cross-check on the diffuse positron fluxes expected from
CBMSPs on Earth (with the same value of $D_{0}=3\times 10^{28}$\,cm$^2$/s).
We reproduce the same shape of the diffuse positron spectrum from those
24~spiders, with  the same break at 300~GeV (due to source 1 in
Table~\ref{table:spiders}, which they placed at an underestimated distance).
Note, however, that the spider spectra calculated by Venter {\it  et al.\/}
miss the low-energy tail visible in Figure~\ref{fig:Fiso} which is caused
by the energy losses of propagating positrons. As a result, their fluxes
at 300\,GeV are about a factor three higher. Additionally, the different 
energy scaling of the diffusion coefficient ($\delta=1/3$ versus $\delta=0.6$)
leads to minor differences in the predicted  fluxes.


Figure~\ref{fig:Fiso} shows the diffuse positron fluxes on Earth
calculated in the isotropic approximation from our source models ABC
(Table~\ref{table:models}), compared to the AMS-02 data.
We show the positron spectra from the currently known population of 52 CBMSPs
with dashed lines (A$'$, B$'$ and C$'$ in Table~\ref{table:models}), and those
of the simulated population of 5000~CBMSPs with dotted lines (A$''$, B$''$
and C$''$; see Section~\ref{sec:spiders-total} for details).
Together, they approximate the total diffuse positron flux on Earth from
spiders in our Galaxy (A, B and C; solid lines in Figure~\ref{fig:Fiso}).
At 100 GeV, our model predicts a total diffuse positron flux between 25 (A)
and 1000 times (C) lower than that measured by AMS-02.
We conclude that the diffuse positron flux that our model predicts from the
total Galactic population of CBMSPs is only a minor contribution to the
observed flux.


We find that the simulated population of 5000~CBMSPs (dotted lines in
Fig.~\ref{fig:Fiso}) contributes with about the same positron flux at
100\,GeV as the currently known population (dashed
lines). Furthermore, the positron flux from spiders after isotropic
diffusion is suppressed above a few TeV. As expected, we find that the
closest spiders dominate the positron flux: the five spiders with the
highest positron flux (three BWs and two RBs) are all within less than
1\,kpc from Earth.
We refer to Appendix~\ref{sec:appendix} for further details on individual
sources.
The contribution from the simulated population is softer, i.e., its high
energy flux is suppressed relative to the one from the currently known
spiders, because of more severe energy losses.


Our results for the diffuse positron flux on Earth with anisotropic transport
due to the ordered Galactic magnetic field are shown in Figure~\ref{fig:Faniso}.
We find that the total fluxes and spectra from CBMSPs are not drastically
modified, but in this case they are clearly dominated by the currently known
population of spiders (dashed lines in Fig.~\ref{fig:Faniso}). The simulated
Galactic population of 5000~CBMSPs contributes with less than 10\% of the
flux at 100\,GeV in this case.  This can be understood as follows; for the
chosen parameters of $D_\|$ and $D_\perp$, the effective volume filled
by positrons emitted by a single source is reduced by a factor
$\simeq 100$~\cite{Giacinti:2017dgt}. Therefore the number of sources
contributing
to the locally measured CR flux is correspondingly reduced and, as a result,
one expects the local flux even in the 100\,GeV range to be dominated by few
local sources~\cite{Kachelriess:2015oua,Kachelriess:2017yzq}.
Correspondingly, the flux of sources in the simulated population---which
are at larger distances---is suppressed. Note, however, that the uncertainties
in current models of the Galactic magnetic field~\cite{Boulanger:2018zrk} are
too large to identify the dominating sources.


Finally, we consider the flux from two nearby and yet undiscovered spiders
near the Galactic plane. In Section~\ref{sec:spiders-total}, we have
estimated that 2--3 such spiders may be "hidden" in the Solar neighborhood
and near the Galactic plane ($|\ell|< 5^\circ$ and $d<1$\,kpc). 
We show in Figure~\ref{fig:Faniso} their contribution to the local diffuse
positron flux, assuming that their line of sight to Earth is either parallel
or perpendicular to the ordered Galactic magnetic field.
The first case (labelled D) represents the effects of anisotropic diffusion in a favorable orientation (along magnetic field lines) while 
the second case (E) shows the least favorable diffusion direction
(across magnetic field lines).
The difference in flux is drastic, three orders of magnitude at 100\,GeV
(Figure~\ref{fig:Faniso}),  although the assumed distance (0.5\,kpc) and
injected spectrum (average of A) are exactly the same.

\begin{figure}[h]
\centering
\includegraphics[width=.8\textwidth,angle=0]{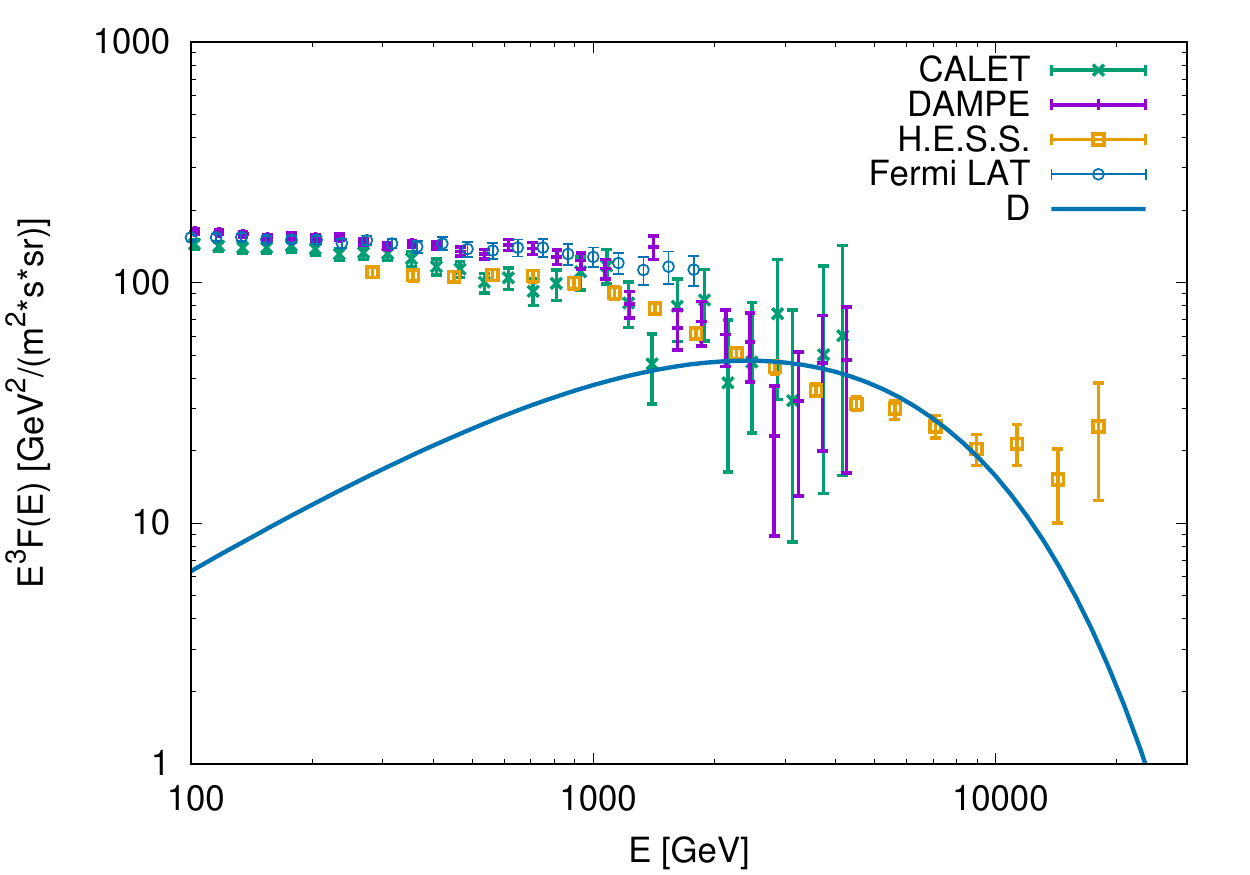}
\caption{\label{HE} Measurements of the diffuse flux of electrons and
  positrons on Earth from CALET, DAMPE, Fermi-LAT and H.E.S.S. compared
  to the flux from our case D.}
\end{figure}

The positron spectrum from our model D reaches the diffuse flux measured
by AMS-02  at 600\,GeV. Taken at face value, this case predicts a second peak
in the positron flux at 2--6\,TeV. In Fig.~\ref{HE}, we compare additionally
the combined electron and positron flux from the same source to measurements
from CALET~\cite{Adriani:2018ktz}, DAMPE~\cite{Ambrosi:2017wek},
Fermi-LAT~\cite{Abdollahi:2017nat} and H.E.S.S.~\cite{Abdalla:2017brm};
the predicted flux from such a source agrees with the existing observations
in the 1--10\,TeV range. Note, however, that the result of case~D might be
overly optimistic; since the Sun is located inside the Local Bubble, the
magnetic field in the bubble wall will act as a shield, reducing the CR flux.
In Ref.~\cite{Bouyahiaoui:2018lew}, it was shown that this reduction is
rather severe in the case of a young source like Vela. For continuous sources
like spiders, the flux suppression due to the Local Bubble should be
smaller. Detailed calculations of positron propagation
taking into account the local magnetic field structure would be needed
to quantify properly this effect.
We conclude that nearby spiders close to the Galactic plane may give a
substantial contribution both to the observed flux of positrons above a
few hundred GeV  and the combined flux of electrons and positrons
above TeV energies, if their line-of-sight to Earth is aligned with the
ordered Galactic magnetic field.

\section{Summary and conclusions}
\label{sec:summary}

We have determined the Galactic scale height of spiders as
$z_\e=0.4\pm0.1$\,kpc, and found that the observed population of spiders is
strongly biased against  small Galactic heights, $|z| < 0.5$\,kpc
(Section~\ref{sec:spiders-current}).
We attribute this
to selection effects of previous searches, which have avoided low 
Galactic latitudes to minimize gamma-ray background and interstellar absorption.
Increasing the sensitivity of CBMSP searches (e.g. with the latest Fermi-LAT source catalog) 
and pushing them to lower Galactic latitudes ($|\ell|\lsim 1^\circ-5^\circ$) will allow us 
to find more nearby spiders, some of which may be
particularly important in terms of cosmic-ray positrons.

We have presented in Section~\ref{sec:pairs} a simple physical model for the fluxes and 
spectra of ``tertiary" pairs, after being reaccelerated at the intrabinary shock in CBMSPs.
We have found that the normalization of such tertiary pair spectra depends mainly on the input rate 
as well as the minimum energy of the pairs ($K\simeq\dot{N}'_\mathrm{p}~E_{\min}$). 
Taking $E_{\min}$ and $\dot N$ from current models, we find that each spider injects between 
$10^{32}$ and a few times $10^{34}$ pairs/s into the interstellar medium, with energies up to 10~TeV.
We also pointed out that, no matter how much kinetic energy is available at the
shock for reacceleration, after taking into account synchrotron losses
the energy output in pairs is limited to 10\% of $L_\mathrm{sd}$. 

Finally, we found in Section~\ref{sec:results} that the contribution from spiders to the diffuse positron 
flux on Earth is less than 5\% of the diffuse flux measured by AMS-02.
As expected, the five spiders that produce the highest positron fluxes are all within 1 kpc (App.~\ref{sec:appendix}).
We also find that the effects of anisotropic diffusion modify strongly the
contribution from individual spiders.
Accurate distances to the nearest sources (e.g. from Gaia's final data) and good knowledge of the
local structure of the Galactic magnetic field will narrow down the predicted range of positron fluxes. 
In any case, even with the moderate 10\% efficiency from our source injection model, we conclude
that one single nearby CBMSP at 0.5\,kpc on the Galactic plane can contribute
significantly to the positron flux at energies above a few hundred GeV
and to the combined electron and positron flux above $\simeq$\,1~TeV,
if it is favorably positioned with respect to the ordered Galactic field.

\appendix
\section{Injected pair spectra}
\label{sec:appendix}

\begin{figure}[h]
\centering
\includegraphics[width=.7\textwidth,angle=-90]{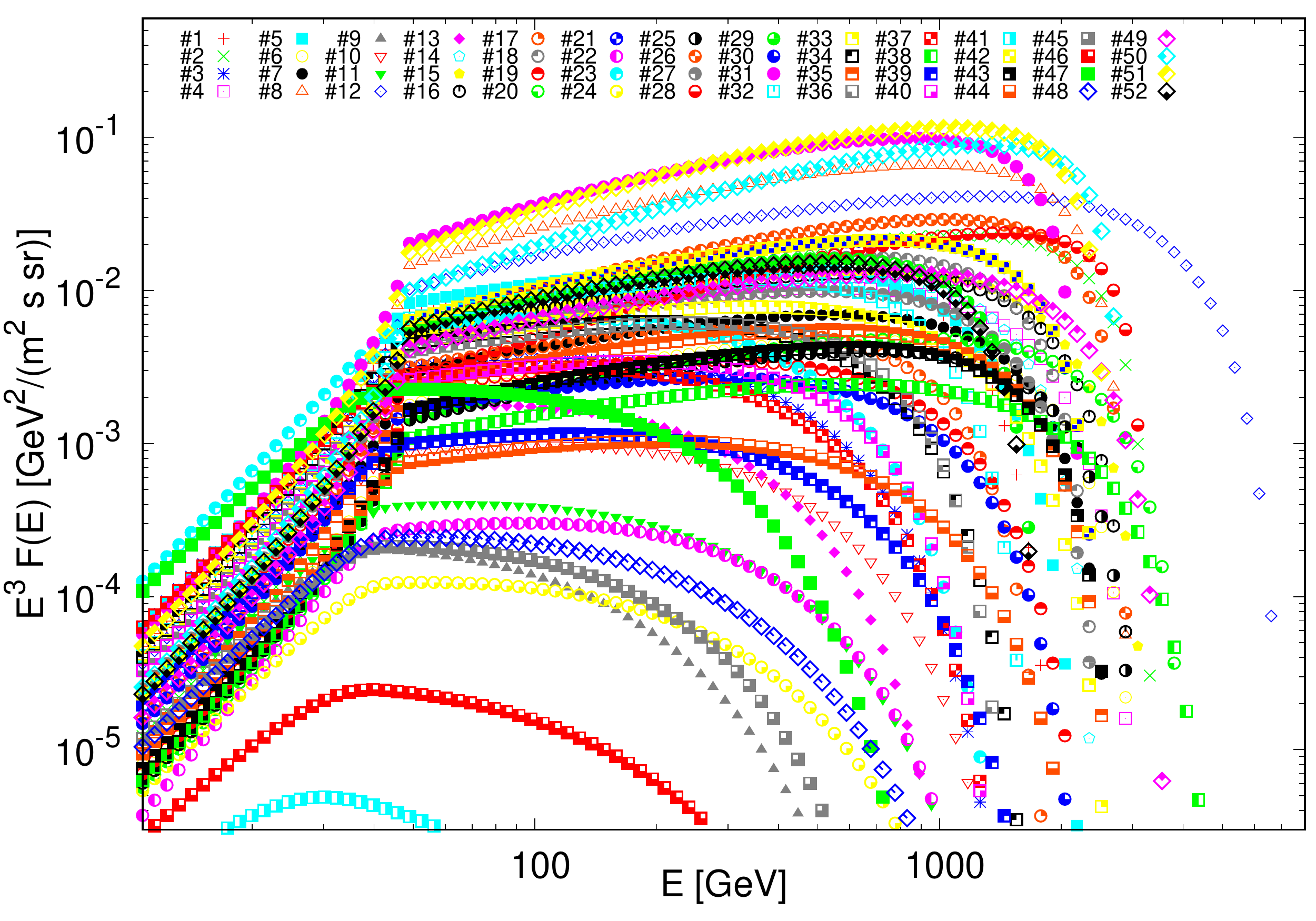}
\caption{\label{fig:e+} Diffuse positron fluxes on Earth for all 52 known CBMSPs, from our model A.}
\end{figure}

We give in this Appendix the model parameters for our three models
(Tables~\ref{table:modelA}, \ref{table:modelB} and ~\ref{table:modelC}), applied to each of the 52 known CBMSPs
(Section~\ref{sec:spiders-current}).
Figure~\ref{fig:e+} shows the diffuse positron fluxes from our model A 
for each of the 52 individual CBMSPs.
The top 5 sources in terms of flux at 1~TeV are source 31, 51, 50, 8 
and 12 in Table~\ref{table:spiders}, at distances of 0.7, 0.6, 0.5, 0.75 and 0.9 kpc, respectively.
The two faintest sources are 46 and 41, at 4.6 and 6.1 kpc from Earth, respectively.
The "bump" at 3 TeV in our isotropic diffuse spectra (models A and B in Figure~\ref{fig:Fiso}) is due to source 12, which has 
the highest $E_\mathrm{top}$ (Figure~\ref{fig:e+} and Tables~\ref{table:modelA} and \ref{table:modelB}).
At energies lower than E$_{\min}$ the diffuse positron spectra show an artificial sharp decrease in flux, which is due to our assumed step-like low energy cut-off in the injected spectra.

\input{model_A.tex}\label{table:modelA}

\input{model_B.tex}\label{table:modelB}

\input{model_C.tex}\label{table:modelC}

\acknowledgments
We thank C. Venter and A. Harding for kindly sharing the details and
code of their work.
We acknowledge the support of the PHAROS COST Action (CA16214).
This work was supported by the Spanish Ministry of Science, Innovation 
and Universities through grant CAS19/00228.



\bibliography{biblio_cr}

\end{document}

%% file: source_table_clean.tex
\begin{table}
  \tiny
\begin{tabular}{|lll|ccccccccr|}
\hline
ID [V15] & Name & type & P$_b$ & a/$\mathrm{10^{10}}$ & M$_{c,min}$ & P$_s$ & B$_s$/$\mathrm{10^{8}}$ & $L_{sd}$/$\mathrm{10^{34}}$ & d & x y z & Ref. \\
-  & - & - & $\mathrm{h}$ & cm & $\mathrm{M_{\odot}}$ & $\mathrm{ms}$  & G & $\mathrm{erg\,s^{-1}}$ & $\mathrm{kpc}$ & $\mathrm{kpc}$  & - \\
\hline
 1	 [18]	 &	 J1023+0038	 &	 RB	 &	 4.8	& 13   &	 0.2	 &	 1.7	 &	 2.2	     &	 5.7	 &	 1.4[0.6]&	 0.43		 0.85		 0.98	 &	 \citep{Archibald09}	 \\	 
 2	 [0]	 &	 J1048+2339	 &	 RB	 &	 6	& 15   &	 0.3	 &	 4.7	 &	 7.6	     &	 1.2	 &	 0.7	 &	 0.27		 0.18		 0.62	 &	 \citep{Deneva16}	 \\	 
 3	 [0]	 &	 J1227-4853	 &	 RB	 &	 6.9	& 16   &	 0.14	 &	 1.7	 &	 2.8	     &	 9.1	 &	 2.5	 &	 -1.2		 2.1		 0.6	 &	 \citep{Roy15}	 \\	 
 4	 [0]	 &	 J1306-40	 &	 RB	 &	 26	& -    &	 -	 &	 2.2	 &	 -	     &	 -	 &	 1.2	 &	 -0.65		 0.9		 0.45	 &	 \citep{Keane17}	 \\	 
 5	 [0]	 &	 J1431-4715	 &	 RB	 &	 11	& 22   &	 0.12	 &	 2	 &	 3.4	     &	 6.8	 &	 1.5	 &	 -1.1		 0.94		 0.32	 &	 \citep{Bates15}	 \\	 
 6	 [0]	 &	 J1622-0315	 &	 RB	 &	 3.9	& 11   &	 0.1	 &	 3.9	 &	 4.3	     &	 0.77	 &	 1.1	 &	 -0.93		 -0.18		 0.56	 &	 \citep{Sanpa16}	 \\	 
 7	 [19]	 &	 J1628-3205	 &	 RB	 &	 5	& 13   &	 0.16	 &	 3.2	 &	 -	     &	 1.4	 &	 1.2	 &	 -1.1		 0.26		 0.24	 &	 \citep{Ray12}	 \\	 
 8	 [20]	 &	 J1723-2837	 &	 RB	 &	 15	& 27   &	 0.24	 &	 1.9	 &	 2.4	     &	 4.7	 &	 0.75	 &	 -0.75		 0.031		 0.056	 &	 \citep{Crawford13}	 \\	 
 9	 [21]	 &	 J1816+4510	 &	 RB	 &	 8.7	& 19   &	 0.16	 &	 3.2	 &	 7.5	     &	 5.2	 &	 4.5[2.4]&	 -1.2		 -3.9		 1.9	 &	 \citep{Kaplan12}	 \\	 
 10	 [0]	 &	 J1908+2105	 &	 RB	 &	 3.5	& 10   &	 0.055	 &	 2.6	 &	 -	     &	 -	 &	 2.6	 &	 -1.5		 -2.1		 0.26	 &	 \citep{Cromartie16}	 \\	 
 11	 [0]	 &	 J1957+2516	 &	 RB	 &	 5.7	& 14   &	 0.1	 &	 4	 &	 6.7	     &	 1.7	 &	 3.1	 &	 -1.4		 -2.8		 -0.11	 &	 \citep{Stovall16}	 \\	 
 12	 [22]	 &	 J2129-0429	 &	 RB	 &	 15	& 28   &	 0.37	 &	 7.6	 &	 -	     &	 3.9	 &	 0.9	 &	 -0.47		 -0.54		 -0.54	 &	 \citep{Roberts11}	 \\	 
 13	 [23]	 &	 J2215+5135	 &	 RB	 &	 4.1	& 11   &	 0.22	 &	 2.6	 &	 6	     &	 7.4	 &	 3	 &	 0.51		 -2.9		 -0.22	 &	 \citep{Roberts11}	 \\	 
 14	 [24]	 &	 J2339-0533	 &	 RB	 &	 4.6	& 12   &	 0.26	 &	 2.9	 &	 4.1	     &	 2.3	 &	 1.1[0.4]&	 -0.076		 -0.5		 -0.98	 &	 \citep{Romani11}	 \\	 
 15	 [0]	 &	 J0212+5320	 &	 RBc	 &	 21	& 34   &	 0.3	 &	 -	 &	 -  	     &	 -	 &	 1.1	 &	 0.77		 -0.77		 -0.15	 &	 \citep{Linares17}	 \\	 
 16	 [0]	 &	 J0523-2529	 &	 RBc	 &	 16	& 32   &	 0.8	 &	 -	 &	 -  	     &	 -	 &	 1.1	 &	 0.64		 0.71		 -0.55	 &	 \citep{Strader14}	 \\	 
 17	 [0]	 &	 J0744-2523	 &	 RBc	 &	 2.8	& -    &	 -	 &	 -	 &	 -  	     &	 -	 &	 1.5	 &	 0.72		 1.3		 -0.018	 &	 \citep{Salvetti17}	 \\	 
 18	 [0]	 &	 J0838.8-2829	 &	 RBc	 &	 5.1	& -    &	 -	 &	 -	 &	 -  	     &	 -	 &	 1	 &	 0.33		 0.93		 0.14	 &	 \citep{Rea17}	 \\	 
 19	 [0]	 &	 J0954.8-3948	 &	 RBc	 &	 9.3	& -    &	 -	 &	 -	 &	 -  	     &	 -	 &	 1.7	 &	 0.002		 1.7		 0.34	 &	 \citep{Li18}	 \\	 
 20	 [0]	 &	 J1302-3258	 &	 RBc	 &	 15	& 27   &	 0.15	 &	 3.8	 &	 -  	     &	 0.5	 &	 0.96	 &	 -0.48		 0.68		 0.48	 &	 \citep{Hessels11}	 \\	 
 21	 [0]	 &	 J2039-5618	 &	 RB	 &	 5.4	& -    &	 -	 &	 2.6	 &	 -  	     &	 -	 &	 0.9	 &	 -0.68		 0.23		 -0.54	 &	 \citep{Salvetti15}	 \\	 
 22	 [0]	 &	 J2333.1-5527	 &	 RBc	 &	 6.9	& -    &	 -	 &	 -	 &	 -  	     &	 -	 &	 3.1	 &	 -1.3		 0.95		 -2.6	 &	 \citep{Swihart20}	 \\
 \hline
 23	 [11]	 &	 B1957+20	 &	 BW	 &	 9.2	& 19   &	 0.021	 &	 1.6	 &	 3.3	     &	 16	 &	 2.5[1.5]&	 -1.3		 -2.1		 -0.2	 &	 \citep{Fruchter88}	 \\	 
 24	 [2]	 &	 J0610-2100	 &	 BW	 &	 6.9	& 16   &	 0.025	 &	 3.9	 &	 4.4	     &	 0.85	 &	 3.5	 &	 2.2		 2.5		 -1.1	 &	 \citep{Burgay06}	 \\	 
 25	 [13]	 &	 J2051-0827	 &	 BW	 &	 2.4	& 7.8  &	 0.027	 &	 4.5	 &	 4.9	     &	 0.55	 &	 1	 &	 -0.7		 -0.57		 -0.53	 &	 \citep{Stappers96}	 \\	 
 26	 [1]	 &	 J0023+0923	 &	 BW	 &	 3.3	& 9.6  &	 0.016	 &	 3	 &	 3.8	     &	 1.6	 &	 0.7	 &	 0.15		 -0.39		 -0.56	 &	 \citep{Hessels11}	 \\	 
 27	 [0]	 &	 J0251+2606	 &	 BW	 &	 4.9	& 13   &	 0.024	 &	 2.5	 &	 -  	     &	 -	 &	 1.2	 &	 0.94		 -0.46		 -0.59	 &	 \citep{Cromartie16}	 \\	 
 28	 [0]	 &	 J0636+5128	 &	 BW	 &	 1.6	& 5.9  &	 0.007	 &	 2.9	 &	 2	     &	 0.56	 &	 0.5	 &	 0.46		 -0.13		 0.16	 &	 \citep{Stovall14}	 \\	 
 29	 [0]	 &	 J0952-0607	 &	 BW	 &	 6.4	& 15   &	 0.02	 &	 1.4	 &	 -  	     &	 -	 &	 0.97	 &	 0.35		 0.71		 0.56	 &	 \citep{Bassa17}	 \\	 
 30	 [3]	 &	 J1124-3653	 &	 BW	 &	 5.4	& 13   &	 0.027	 &	 2.4	 &	 -  	     &	 1.6	 &	 1.7	 &	 -0.38		 1.5		 0.66	 &	 \citep{Hessels11}	 \\	 
 31	 [4]	 &	 J1301+0833	 &	 BW	 &	 6.5	& 15   &	 0.024	 &	 1.8	 &	 -  	     &	 6.7	 &	 0.7	 &	 -0.15		 0.17		 0.66	 &	 \citep{Ray12}	 \\	 
 32	 [5]	 &	 J1311-3430	 &	 BW	 &	 1.6	& 5.8  &	 0.008	 &	 2.6	 &	 4.7	     &	 4.9	 &	 1.4	 &	 -0.75		 0.98		 0.66	 &	 \citep{Romani12}	 \\	 
 33	 [6]	 &	 J1446-4701	 &	 BW	 &	 6.7	& 15   &	 0.0019	 &	 2.2	 &	 3	     &	 3.7	 &	 1.5	 &	 -1.2		 0.9		 0.3	 &	 \citep{Keith12}	 \\	 
 34	 [0]	 &	 J1513-2550	 &	 BW	 &	 4.3	& 11   &	 0.02	 &	 2.1	 &	 4.3	     &	 8.8	 &	 2	 &	 -0.82		 -0.66		 1.7	 &	 \citep{Sanpa16}	 \\	 
 35	 [7]	 &	 J1544+4937	 &	 BW	 &	 2.8	& 8.6  &	 0.018	 &	 2.2	 &	 1.6	     &	 1.2	 &	 1.2	 &	 -0.14		 -0.75		 0.92	 &	 \citep{Bhat13}	 \\	 
 36	 [0]	 &	 J1641+8049	 &	 BW	 &	 2.2	& 7.4  &	 0.04	 &	 2	 &	 -	     &	 4.3	 &	 1.7	 &	 0.58		 -1.3		 0.89	 &	 \citep{Lynch18}	 \\	 
 37	 [8]	 &	 J1731-1847	 &	 BW	 &	 7.5	& 17   &	 0.04	 &	 2.3	 &	 4.9	     &	 7.8	 &	 2.5	 &	 -2.5		 -0.3		 0.35	 &	 \citep{Bates11}	 \\	 
 38	 [9]	 &	 J1745+1017	 &	 BW	 &	 18	& 29   &	 0.014	 &	 2.6	 &	 1.7	     &	 0.58	 &	 1.3[1.4]&	 -1.0		 -0.7		 0.43	 &	 \citep{Barr13}	 \\	 
 39	 [0]	 &	 J1805+06	 &	 BW	 &	 8.1	& 17   &	 0.023	 &	 2.1	 &	 -	     &	 -	 &	 2.5	 &	 -2		 -1.3		 0.56	 &	 \citep{Cromartie16}	 \\	 
 40	 [10]	 &	 J1810+1744	 &	 BW	 &	 3.6	& 10   &	 0.044	 &	 1.7	 &	 -	     &	 3.9	 &	 2	 &	 -1.4		 -1.3		 0.58	 &	 \citep{Hessels11}	 \\	 
 41	 [0]	 &	 J1928]+1245	 &	 BW	 &	 3.3	& 9.6  &	 0.009	 &	 3	 &	 1.7	     &	 0.24	 &	 6.1	 &	 -4		 -4.6		 -0.24	 &	 \citep{Parent19}	 \\	 
 42	 [0]	 &	 J1946-5403	 &	 BW	 &	 3.1	& 9.2  &	 0.021	 &	 2.7	 &	 -	     &	 -	 &	 0.9	 &	 -0.75		 0.22		 -0.44	 &	 \citep{Camilo15}	 \\	 
 43	 [0]	 &	 J2017-1614	 &	 BW	 &	 2.3	& 7.6  &	 0.03	 &	 2.3	 &	 1.5	     &	 0.7	 &	 1.1	 &	 -0.88		 -0.45		 -0.49	 &	 \citep{Sanpa16}	 \\	 
 44	 [12]	 &	 J2047+1053	 &	 BW	 &	 3	& 9    &	 0.035	 &	 4.3	 &	 -	     &	 1	 &	 2	 &	 -1.0		 -1.6		 -0.67	 &	 \citep{Ray12}	 \\	 
 45	 [0]	 &	 J2052+1218	 &	 BW	 &	 2.6	& 8.2  &	 0.033	 &	 2	 &	 -	     &	 -	 &	 3.9	 &	 -1.9		 -3.1		 -1.3	 &	 \citep{Cromartie16}	 \\	 
 46	 [0]	 &	 J2055+3829	 &	 BW	 &	 3.1	& 9.2  &	 0.023	 &	 2.1	 &	 0.93	     &	 0.36	 &	 4.6	 &	 -0.75		 -4.5		 -0.34	 &	 \citep{Guillemot19}	 \\	 
 47	 [0]	 &	 J2115+5448	 &	 BW	 &	 3.2	& 9.4  &	 0.02	 &	 2.6	 &	 9	     &	 16	 &	 3.4	 &	 0.3		 -3.4		 0.24	 &	 \citep{Sanpa16}	 \\	 
 48	 [14]	 &	 J2214+3000	 &	 BW	 &	 10	& 20   &	 0.014	 &	 3.1	 &	 4.3	     &	 1.9	 &	 3.6[1.3]&	 -0.18		 -3.3		 -1.3	 &	 \citep{Ransom11}	 \\	 
 49	 [15]	 &	 J2234+0944	 &	 BW	 &	 10	& 20   &	 0.015	 &	 3.6	 &	 5.5	     &	 1.7	 &	 1	 &	 -0.18		 -0.74		 -0.65	 &	 \citep{Keith12}	 \\	 
 50	 [16]	 &	 J2241-5236	 &	 BW	 &	 3.4	& 9.8  &	 0.012	 &	 2.2	 &	 2.4	     &	 2.5	 &	 0.5	 &	 -0.27		 0.11		 -0.41	 &	 \citep{Keith11}	 \\	 
 51	 [17]	 &	 J2256-1024	 &	 BW	 &	 5.1	& 13   &	 0.034	 &	 2.3	 &	 -	     &	 5.2	 &	 0.6	 &	 -0.16		 -0.27		 -0.51	 &	 \citep{Boyles11}	 \\	 
 52	 [0]	 &	 J1653-0159	 &	 BWc	 &	 1.2	& -    &	 -	 &	 -	 &	 -	     &	 -	 &	 1	 &	 -0.87		 -0.26		 0.42	 &	 \citep{Romani14}	 \\	 
 \hline
\end{tabular}
\caption{\label{table:spiders} Names and properties of the 52
  currently known CBMSPs: orbital period P$_b$ and separation a,
  minimum mass of the companion star M$_{c,min}$, pulsar spin period
  P$_s$, magnetic field B$_s$ and spin-down luminosity L$_{sd}$, as
  well as distance and Cartesian coordinates. The first column,
  between square brackets, shows the ID number of the systems studied
  by Venter et al. \citep{Venter15} (those with [0] were not
  included/known at the time). "RB" and "BW" are redback and black
  widow sub-types, respectively, and a "c" denotes those candidate
  systems with no published detection of pulsations. When these differ
  from ours, we show between square brackets the distances used by
  Venter et al. \citep{Venter15}. Median values are: P=2.6~ms,
  L$_{sd}$=2.3$\times$10$^{34}$~erg~s$^{-1}$,
  $\dot{P}$=1.3$\times$10$^{-20}$, B$_s$=3.6$\times$10$^{8}$~G,
  P$_b$=5.1~hr, M$_{c,min}$=0.16~M$_\odot$ (RBs), 0.021~M$_\odot$
  (BWs), a=1.3$\times$10$^{11}$~cm. Note: L$_{sd}$ values listed as
  reported (not always corrected for Shklovskii effect, often assume
  I=10$^{45}$~g~cm$^2$). M$_{c,min}$ values for an edge-on orbit
  listed as reported (often assume M=1.4~M$_\odot$ for the pulsar).}
\end{table}

%% file: model_A.tex
\begin{table}
\centering
  \small
\begin{tabular}{|ccccccccc|}
\hline    
ID & R$_\mathrm{sh}$/a & B$_\mathrm{sh}$ & E$_\mathrm{top}$ & $\Omega_2$ & L$_\mathrm{w}$/10$^{32}$ &  K/10$^{32}$ & $L'_\mathrm{p}$/10$^{32}$ & $\dot{N}_\mathrm{p}$/10$^{33}$ \\
- & - & $\mathrm{G}$ & $\mathrm{TeV}$ & - & $\mathrm{erg\,s^{-1}}$ & $\mathrm{erg\,s^{-1}}$ & $\mathrm{erg\,s^{-1}}$ & s$^{-1}$ \\
\hline
1 & 0.27 & 132 & 2.3 & 0.018 & 3.2 & 8.2 & 27 & 10 \\
2 & 0.25 & 54 & 3.6 & 0.016 & 0.85 & 2 & 7.4 & 2.5 \\
3 & 0.26 & 135 & 2.3 & 0.018 & 4.1 & 12 & 41 & 16 \\
4 & 0.22 & 47 & 3.8 & 0.012 & 1.2 & 3.7 & 14 & 4.6 \\
5 & 0.25 & 91 & 2.8 & 0.016 & 2.8 & 9.6 & 33 & 12 \\
6 & 0.29 & 53 & 3.6 & 0.022 & 0.29 & 1.3 & 5 & 1.6 \\
7 & 0.27 & 66 & 3.2 & 0.019 & 0.66 & 2.2 & 8 & 2.7 \\
8 & 0.23 & 66 & 3.2 & 0.013 & 2.9 & 6.9 & 25 & 8.6 \\
9 & 0.25 & 91 & 2.8 & 0.016 & 2.6 & 7.5 & 26 & 9.4 \\
10 & 0.31 & 104 & 2.6 & 0.025 & 0.62 & 3.7 & 13 & 4.6 \\
11 & 0.28 & 64 & 3.3 & 0.02 & 0.63 & 2.7 & 9.9 & 3.4 \\
12 & 0.22 & 10 & 8.2 & 0.012 & 3.1 & 5.8 & 27 & 7.2 \\
13 & 0.27 & 166 & 2 & 0.019 & 4.4 & 10 & 33 & 13 \\
14 & 0.26 & 88 & 2.8 & 0.017 & 1.5 & 3.6 & 12 & 4.5 \\
15 & 0.21 & 44 & 3.9 & 0.012 & 1.7 & 3.7 & 14 & 4.6 \\
16 & 0.20 & 51 & 3.7 & 0.01 & 2.9 & 3.7 & 14 & 4.6 \\
17 & 0.29 & 127 & 2.3 & 0.022 & 1.2 & 3.7 & 12 & 4.6 \\
18 & 0.27 & 92 & 2.7 & 0.019 & 1.2 & 3.7 & 13 & 4.6 \\
19 & 0.25 & 66 & 3.2 & 0.016 & 1.2 & 3.7 & 14 & 4.6 \\
20 & 0.24 & 28 & 5 & 0.014 & 0.24 & 0.89 & 3.6 & 1.1 \\
21 & 0.27 & 92 & 2.7 & 0.018 & 1.2 & 3.9 & 13 & 4.8 \\
22 & 0.26 & 78 & 3 & 0.017 & 1.2 & 3.7 & 13 & 4.6 \\
23 & 0.31 & 129 & 2.3 & 0.024 & 2.3 & 21 & 69 & 26 \\
24 & 0.31 & 36 & 4.4 & 0.025 & 0.13 & 1.4 & 5.7 & 1.8 \\
25 & 0.35 & 52 & 3.7 & 0.031 & 0.091 & 0.97 & 3.7 & 1.2 \\
26 & 0.35 & 70 & 3.1 & 0.032 & 0.19 & 2.6 & 9.3 & 3.2 \\
27 & 0.33 & 81 & 2.9 & 0.027 & 0.37 & 3.7 & 13 & 4.6 \\
28 & 0.41 & 58 & 3.4 & 0.044 & 0.039 & 0.99 & 3.7 & 1.2 \\
29 & 0.32 & 166 & 2 & 0.027 & 0.33 & 3.7 & 12 & 4.6 \\
30 & 0.32 & 84 & 2.9 & 0.026 & 0.26 & 2.6 & 9 & 3.2 \\
31 & 0.32 & 112 & 2.5 & 0.026 & 1 & 9.5 & 32 & 12 \\
32 & 0.41 & 177 & 2 & 0.044 & 0.37 & 7.1 & 22 & 8.9 \\
33 & 0.40 & 58 & 3.4 & 0.043 & 0.11 & 5.5 & 20 & 6.9 \\
34 & 0.34 & 147 & 2.2 & 0.029 & 1.2 & 12 & 39 & 15 \\
35 & 0.36 & 66 & 3.2 & 0.033 & 0.15 & 2 & 7.2 & 2.5 \\
36 & 0.34 & 187 & 1.9 & 0.03 & 0.91 & 6.3 & 20 & 7.9 \\
37 & 0.29 & 108 & 2.5 & 0.022 & 1.7 & 11 & 37 & 14 \\
38 & 0.30 & 17 & 6.5 & 0.023 & 0.063 & 1 & 4.4 & 1.3 \\
39 & 0.31 & 80 & 2.9 & 0.024 & 0.36 & 3.7 & 13 & 4.6 \\
40 & 0.32 & 192 & 1.9 & 0.026 & 0.88 & 5.8 & 18 & 7.2 \\
41 & 0.37 & 31 & 4.7 & 0.036 & 0.02 & 0.46 & 1.8 & 0.57 \\
42 & 0.35 & 94 & 2.7 & 0.031 & 0.34 & 3.7 & 13 & 4.6 \\
43 & 0.35 & 64 & 3.3 & 0.031 & 0.12 & 1.2 & 4.4 & 1.5 \\
44 & 0.33 & 50 & 3.7 & 0.028 & 0.19 & 1.7 & 6.3 & 2.1 \\
45 & 0.34 & 171 & 2 & 0.03 & 0.45 & 3.7 & 12 & 4.6 \\
46 & 0.34 & 39 & 4.2 & 0.031 & 0.054 & 0.66 & 2.6 & 0.83 \\
47 & 0.35 & 237 & 1.7 & 0.031 & 2.3 & 21 & 64 & 27 \\
48 & 0.32 & 41 & 4.1 & 0.026 & 0.21 & 3 & 12 & 3.8 \\
49 & 0.31 & 38 & 4.2 & 0.025 & 0.19 & 2.7 & 11 & 3.4 \\
50 & 0.36 & 84 & 2.9 & 0.034 & 0.25 & 3.9 & 14 & 4.8 \\
51 & 0.31 & 96 & 2.7 & 0.025 & 0.99 & 7.5 & 26 & 9.4 \\
52 & 0.38 & 174 & 2 & 0.038 & 0.34 & 3.7 & 12 & 4.6 \\
\hline
\end{tabular}
\caption{\label{table:modelA} Companion wind, intrabinary shock and
  pair spectral parameters for the 52 spiders, from our model A.}
\end{table}

%% file: model_B.tex
\begin{table}
\centering
  \small
\begin{tabular}{|ccccccccc|}
\hline    
ID & R$_\mathrm{sh}$/a & B$_\mathrm{sh}$ & E$_\mathrm{top}$ & $\Omega_2$ & L$_\mathrm{w}$/10$^{32}$ &  K/10$^{32}$ & $L'_\mathrm{p}$/10$^{32}$ & $\dot{N}_\mathrm{p}$/10$^{33}$ \\
- & - & $\mathrm{G}$ & $\mathrm{TeV}$ & - & $\mathrm{erg\,s^{-1}}$ & $\mathrm{erg\,s^{-1}}$ & $\mathrm{erg\,s^{-1}}$ & s$^{-1}$ \\
\hline
1 & 0.37 & 97 & 2.7 & 0.035 & 1.3 & 0.82 & 4.2 & 5.1 \\
2 & 0.35 & 39 & 4.2 & 0.031 & 0.34 & 0.2 & 1.1 & 1.2 \\
3 & 0.36 & 98 & 2.7 & 0.034 & 1.7 & 1.2 & 6.4 & 7.8 \\
4 & 0.31 & 34 & 4.5 & 0.024 & 0.47 & 0.37 & 2.1 & 2.3 \\
5 & 0.35 & 66 & 3.2 & 0.031 & 1.1 & 0.96 & 5.2 & 6 \\
6 & 0.39 & 39 & 4.2 & 0.04 & 0.11 & 0.13 & 0.75 & 0.82 \\
7 & 0.37 & 48 & 3.8 & 0.035 & 0.27 & 0.22 & 1.2 & 1.4 \\
8 & 0.32 & 47 & 3.8 & 0.026 & 1.2 & 0.69 & 3.8 & 4.3 \\
9 & 0.35 & 66 & 3.2 & 0.031 & 1 & 0.75 & 4 & 4.7 \\
10 & 0.42 & 78 & 3 & 0.046 & 0.25 & 0.37 & 2 & 2.3 \\
11 & 0.38 & 47 & 3.8 & 0.037 & 0.25 & 0.27 & 1.5 & 1.7 \\
12 & 0.31 & 7 & 9.7 & 0.024 & 1.2 & 0.58 & 4 & 3.6 \\
13 & 0.37 & 121 & 2.4 & 0.035 & 1.7 & 1 & 5.2 & 6.5 \\
14 & 0.36 & 64 & 3.3 & 0.033 & 0.6 & 0.36 & 1.9 & 2.2 \\
15 & 0.30 & 32 & 4.7 & 0.023 & 0.68 & 0.37 & 2.2 & 2.3 \\
16 & 0.29 & 36 & 4.4 & 0.021 & 1.1 & 0.37 & 2.1 & 2.3 \\
17 & 0.39 & 94 & 2.7 & 0.04 & 0.47 & 0.37 & 1.9 & 2.3 \\
18 & 0.37 & 67 & 3.2 & 0.035 & 0.47 & 0.37 & 2 & 2.3 \\
19 & 0.35 & 48 & 3.8 & 0.031 & 0.47 & 0.37 & 2.1 & 2.3 \\
20 & 0.33 & 20 & 5.8 & 0.028 & 0.095 & 0.089 & 0.55 & 0.56 \\
21 & 0.37 & 67 & 3.2 & 0.035 & 0.49 & 0.39 & 2.1 & 2.4 \\
22 & 0.36 & 57 & 3.5 & 0.033 & 0.47 & 0.37 & 2 & 2.3 \\
23 & 0.41 & 96 & 2.7 & 0.044 & 0.9 & 2.1 & 11 & 13 \\
24 & 0.42 & 27 & 5.1 & 0.046 & 0.054 & 0.14 & 0.86 & 0.9 \\
25 & 0.46 & 39 & 4.2 & 0.056 & 0.036 & 0.097 & 0.55 & 0.61 \\
26 & 0.46 & 53 & 3.6 & 0.057 & 0.076 & 0.26 & 1.4 & 1.6 \\
27 & 0.43 & 61 & 3.4 & 0.049 & 0.15 & 0.37 & 2 & 2.3 \\
28 & 0.53 & 46 & 3.9 & 0.075 & 0.016 & 0.099 & 0.55 & 0.62 \\
29 & 0.43 & 125 & 2.4 & 0.048 & 0.13 & 0.37 & 1.9 & 2.3 \\
30 & 0.42 & 63 & 3.3 & 0.047 & 0.11 & 0.26 & 1.4 & 1.6 \\
31 & 0.42 & 84 & 2.9 & 0.047 & 0.41 & 0.95 & 5 & 5.9 \\
32 & 0.52 & 138 & 2.2 & 0.073 & 0.15 & 0.71 & 3.5 & 4.4 \\
33 & 0.52 & 46 & 3.9 & 0.072 & 0.044 & 0.55 & 3.1 & 3.4 \\
34 & 0.44 & 111 & 2.5 & 0.052 & 0.48 & 1.2 & 6.2 & 7.6 \\
35 & 0.47 & 50 & 3.7 & 0.058 & 0.061 & 0.2 & 1.1 & 1.2 \\
36 & 0.45 & 141 & 2.2 & 0.053 & 0.36 & 0.63 & 3.1 & 3.9 \\
37 & 0.40 & 80 & 2.9 & 0.041 & 0.66 & 1.1 & 5.7 & 6.8 \\
38 & 0.40 & 12 & 7.5 & 0.042 & 0.025 & 0.1 & 0.66 & 0.64 \\
39 & 0.41 & 59 & 3.4 & 0.045 & 0.14 & 0.37 & 2 & 2.3 \\
40 & 0.42 & 144 & 2.2 & 0.047 & 0.35 & 0.58 & 2.8 & 3.6 \\
41 & 0.49 & 24 & 5.4 & 0.063 & 0.0079 & 0.046 & 0.27 & 0.29 \\
42 & 0.46 & 71 & 3.1 & 0.055 & 0.14 & 0.37 & 2 & 2.3 \\
43 & 0.46 & 48 & 3.8 & 0.055 & 0.049 & 0.12 & 0.67 & 0.76 \\
44 & 0.44 & 38 & 4.3 & 0.051 & 0.078 & 0.17 & 0.96 & 1 \\
45 & 0.45 & 130 & 2.3 & 0.053 & 0.18 & 0.37 & 1.9 & 2.3 \\
46 & 0.45 & 30 & 4.8 & 0.054 & 0.022 & 0.066 & 0.39 & 0.41 \\
47 & 0.46 & 180 & 2 & 0.055 & 0.9 & 2.1 & 10 & 13 \\
48 & 0.42 & 31 & 4.7 & 0.047 & 0.083 & 0.3 & 1.8 & 1.9 \\
49 & 0.42 & 29 & 4.9 & 0.046 & 0.077 & 0.27 & 1.6 & 1.7 \\
50 & 0.47 & 64 & 3.3 & 0.06 & 0.099 & 0.39 & 2.1 & 2.4 \\
51 & 0.42 & 72 & 3.1 & 0.046 & 0.4 & 0.75 & 4 & 4.7 \\
52 & 0.50 & 134 & 2.3 & 0.066 & 0.14 & 0.37 & 1.8 & 2.3 \\
\hline
\end{tabular}
\caption{\label{table:modelB} Companion wind, intrabinary shock and
  pair spectral parameters for the 52 spiders, from our model B.}
\end{table}

%% file: model_C.tex
\begin{table}
\centering
  \small
\begin{tabular}{|ccccccccc|}
\hline    
ID & R$_\mathrm{sh}$/a & B$_\mathrm{sh}$ & E$_\mathrm{top}$ & $\Omega_2$ & L$_\mathrm{w}$/10$^{32}$ &  K/10$^{32}$ & $L'_\mathrm{p}$/10$^{32}$ & $\dot{N}_\mathrm{p}$/10$^{33}$ \\
- & - & $\mathrm{G}$ & $\mathrm{TeV}$ & - & $\mathrm{erg\,s^{-1}}$ & $\mathrm{erg\,s^{-1}}$ & $\mathrm{erg\,s^{-1}}$ & s$^{-1}$ \\
\hline
1 & 0.45 & 79 & 1.7 & 0.053 & 0.64 & 0.16 & 0.93 & 2.5 \\
2 & 0.43 & 32 & 2.7 & 0.048 & 0.17 & 0.04 & 0.25 & 0.62 \\
3 & 0.44 & 80 & 1.5 & 0.052 & 0.83 & 0.25 & 1.4 & 3.9 \\
4 & 0.39 & 27 & 2.2 & 0.038 & 0.24 & 0.074 & 0.45 & 1.2 \\
5 & 0.43 & 53 & 1.6 & 0.049 & 0.56 & 0.19 & 1.1 & 3 \\
6 & 0.48 & 32 & 3.2 & 0.061 & 0.057 & 0.026 & 0.17 & 0.41 \\
7 & 0.45 & 39 & 2.6 & 0.054 & 0.13 & 0.044 & 0.28 & 0.69 \\
8 & 0.40 & 38 & 1.8 & 0.041 & 0.58 & 0.14 & 0.79 & 2.1 \\
9 & 0.43 & 53 & 1.7 & 0.048 & 0.51 & 0.15 & 0.86 & 2.3 \\
10 & 0.50 & 64 & 2.2 & 0.068 & 0.12 & 0.074 & 0.45 & 1.2 \\
11 & 0.46 & 38 & 2.5 & 0.057 & 0.13 & 0.054 & 0.34 & 0.85 \\
12 & 0.38 & 6 & 1.9 & 0.038 & 0.62 & 0.12 & 0.67 & 1.8 \\
13 & 0.45 & 99 & 1.6 & 0.054 & 0.87 & 0.21 & 1.2 & 3.2 \\
14 & 0.44 & 52 & 2.2 & 0.051 & 0.3 & 0.072 & 0.43 & 1.1 \\
15 & 0.38 & 25 & 2.2 & 0.036 & 0.34 & 0.074 & 0.45 & 1.2 \\
16 & 0.36 & 28 & 2.2 & 0.033 & 0.57 & 0.074 & 0.45 & 1.2 \\
17 & 0.48 & 77 & 2.2 & 0.061 & 0.24 & 0.074 & 0.45 & 1.2 \\
18 & 0.45 & 55 & 2.2 & 0.054 & 0.24 & 0.074 & 0.45 & 1.2 \\
19 & 0.43 & 39 & 2.2 & 0.048 & 0.24 & 0.074 & 0.45 & 1.2 \\
20 & 0.41 & 16 & 3.7 & 0.043 & 0.047 & 0.018 & 0.12 & 0.28 \\
21 & 0.45 & 55 & 2.2 & 0.053 & 0.25 & 0.077 & 0.46 & 1.2 \\
22 & 0.44 & 46 & 2.2 & 0.051 & 0.24 & 0.074 & 0.45 & 1.2 \\
23 & 0.50 & 80 & 1.2 & 0.066 & 0.45 & 0.42 & 2.2 & 6.5 \\
24 & 0.50 & 22 & 3.1 & 0.068 & 0.027 & 0.029 & 0.19 & 0.45 \\
25 & 0.54 & 33 & 3.6 & 0.081 & 0.018 & 0.019 & 0.13 & 0.3 \\
26 & 0.55 & 45 & 2.5 & 0.083 & 0.038 & 0.051 & 0.32 & 0.8 \\
27 & 0.52 & 51 & 2.2 & 0.073 & 0.074 & 0.074 & 0.45 & 1.2 \\
28 & 0.61 & 39 & 3.5 & 0.1 & 0.0078 & 0.02 & 0.13 & 0.31 \\
29 & 0.51 & 104 & 2.2 & 0.071 & 0.066 & 0.074 & 0.45 & 1.2 \\
30 & 0.51 & 52 & 2.5 & 0.07 & 0.053 & 0.051 & 0.32 & 0.8 \\
31 & 0.51 & 70 & 1.6 & 0.069 & 0.21 & 0.19 & 1.1 & 3 \\
32 & 0.61 & 119 & 1.8 & 0.1 & 0.074 & 0.14 & 0.82 & 2.2 \\
33 & 0.60 & 39 & 1.9 & 0.1 & 0.022 & 0.11 & 0.65 & 1.7 \\
34 & 0.53 & 93 & 1.5 & 0.076 & 0.24 & 0.24 & 1.3 & 3.8 \\
35 & 0.55 & 42 & 2.7 & 0.083 & 0.031 & 0.04 & 0.25 & 0.62 \\
36 & 0.53 & 118 & 1.8 & 0.077 & 0.18 & 0.13 & 0.73 & 2 \\
37 & 0.48 & 66 & 1.5 & 0.062 & 0.33 & 0.22 & 1.2 & 3.4 \\
38 & 0.49 & 10 & 3.5 & 0.063 & 0.013 & 0.02 & 0.14 & 0.32 \\
39 & 0.50 & 49 & 2.2 & 0.067 & 0.072 & 0.074 & 0.45 & 1.2 \\
40 & 0.51 & 120 & 1.9 & 0.07 & 0.18 & 0.12 & 0.67 & 1.8 \\
41 & 0.57 & 20 & 4.7 & 0.09 & 0.0039 & 0.0091 & 0.066 & 0.14 \\
42 & 0.54 & 60 & 2.2 & 0.08 & 0.068 & 0.074 & 0.45 & 1.2 \\
43 & 0.54 & 41 & 3.3 & 0.08 & 0.025 & 0.024 & 0.16 & 0.38 \\
44 & 0.53 & 32 & 2.9 & 0.075 & 0.039 & 0.034 & 0.22 & 0.52 \\
45 & 0.53 & 109 & 2.2 & 0.077 & 0.09 & 0.074 & 0.45 & 1.2 \\
46 & 0.54 & 25 & 4.1 & 0.079 & 0.011 & 0.013 & 0.092 & 0.21 \\
47 & 0.54 & 151 & 1.2 & 0.08 & 0.45 & 0.43 & 2.3 & 6.7 \\
48 & 0.51 & 25 & 2.4 & 0.07 & 0.041 & 0.06 & 0.37 & 0.94 \\
49 & 0.51 & 24 & 2.5 & 0.069 & 0.039 & 0.054 & 0.34 & 0.85 \\
50 & 0.56 & 54 & 2.2 & 0.086 & 0.049 & 0.077 & 0.46 & 1.2 \\
51 & 0.50 & 59 & 1.7 & 0.068 & 0.2 & 0.15 & 0.86 & 2.3 \\
52 & 0.58 & 115 & 2.2 & 0.093 & 0.068 & 0.074 & 0.45 & 1.2 \\
\hline
\end{tabular}
\caption{\label{table:modelC} Companion wind, intrabinary shock and
  pair spectral parameters for the 52 spiders, from our model C.}
\end{table}